\documentclass[12pt,twoside]{article}
\usepackage[mathscr]{eucal}
\usepackage{amsmath,amsfonts,amssymb,amsthm,mathabx,empheq}
\usepackage{times}
\usepackage{pdfsync}
\usepackage{cite}
\usepackage{url}
\usepackage{hyperref}
\usepackage{tensor}
\usepackage{color}
\usepackage{multicol}
\usepackage{bbold}
\usepackage{tikz}
\usetikzlibrary{calc,decorations.markings}

\advance\textheight\topskip
\allowdisplaybreaks[1]

\newcommand\td{\text{d}}

\newcommand{\p}{\partial}

\newcommand{\be}{\begin{equation}}
\newcommand{\ee}{\end{equation}}
\newcommand{\bea}{\begin{eqnarray}}
\newcommand{\eea}{\end{eqnarray}}
\def\nn{\nonumber}
\def\bz{\bar z}

\newcommand*\xbar[1]{%
  \hbox{%
    \vbox{%
      \hrule height 0.5pt 
      \kern0.3ex
      \hbox{%
        \kern-0.0em
        \ensuremath{#1}%
        \kern-0.0em
      }%
    }%
  }%
}
\def\bra#1{\left\langle #1\right|}
\def\ket#1{\left| #1\right\rangle}
\def\>{\rangle} 
\def\<{\langle} 

\DeclareFontFamily{OT1}{rsfs}{} \DeclareFontShape{OT1}{rsfs}{m}{n}{
<-7> rsfs5 <7-10> rsfs7 <10-> rsfs10}{}
\DeclareMathAlphabet{\mycal}{OT1}{rsfs}{m}{n}

\begin{document}
\title{T\={T} deformed soft theorem}

\author{Song He, Pujian Mao and Xin-Cheng Mao}

\date{}

\def\mytitle{T\={T} deformed soft theorem}

\addtolength{\headsep}{4pt}

\begin{centering}

  \vspace{1cm}

  \textbf{\Large{\mytitle}}

  \vspace{1.5cm}

  {\large Song He$^{{\cal T}_{zz},{\cal T}_{z\bz}}$, Pujian Mao$^{{\cal T}_{\bz \bz}}$ and Xin-Cheng Mao$^{{\cal T}_{zz}}$ }

\vspace{0.5cm}

\begin{minipage}{.9\textwidth}\small \it  \begin{center}
    ${}^{{\cal T}_{zz}}$ Center for Theoretical Physics and College of Physics,\\ Jilin University, 2699 Qianjin Street,
   Changchun 130012, China
 \end{center}
\end{minipage}

\vspace{0.3cm}

\begin{minipage}{.9\textwidth}\small \it  \begin{center}
     ${}^{{\cal T}_{\bz \bz}}$ Center for Joint Quantum Studies and Department of Physics,\\
     School of Science, Tianjin University, 135 Yaguan Road, Tianjin 300350, China
 \end{center}
\end{minipage}

\vspace{0.3cm}

\begin{minipage}{.9\textwidth}\small \it  \begin{center}
     ${}^{{\cal T}_{z\bz}}$ Max Planck Institute for Gravitational Physics (Albert Einstein Institute),\\
     Am M\"{u}hlenberg 1, 14476 Golm, Germany
 \end{center}
\end{minipage}

\end{centering}


\vspace{1cm}

\begin{center}
\begin{minipage}{.9\textwidth}
  \textsc{Abstract}. In this paper, we derive a $T\bar{T}$ deformed soft graviton theorem in the context of celestial holography. As a concrete example, it illustrates that a two-dimensional irrelevant deformation can be applied to a four-dimensional theory at the level of amplitudes. We argue that the $T\bar{T}$ deformation has a close relation to the loop-correction from the amplitude side which could provide an alternative way to construct an ultraviolet complete quantum theory of gravity.
 \end{minipage}
\end{center}
\thispagestyle{empty}


\section{Introduction}

Newton's constant has a negative mass dimension; therefore, General Relativity is not renormalizable in the usual sense, which 't Hooft conclusively confirms and Veltman in the early seventies \cite{tHooft:1974toh}. As an effective theory, General Relativity can properly describe gravitational interaction at the low energy scale in the Wilson scheme. Then the search for a consistent ultraviolet (UV) completion for General Relativity has been a tremendous physical problem for more than half a century, see, e.g., \cite{Hamber:2009zz} for a comprehensive introduction.

In general, conformal field theory (CFT) is central to describing the fixed points of the renormalization group flow. A common way of flowing away from fixed points to probe the dynamics at higher energy scales is to consider irrelevant deformations of the theory. In particular, it was recently discovered that the composite operator $T\bar{T}$ could lead to a tractable and even solvable irrelevant deformation in two-dimensional (2D) spacetime \cite{Smirnov:2016lqw, Cavaglia:2016oda}. The deformed theories in the deep UV are expected to be UV complete. One crucial piece of evidence is that the $T\bar{T}$ deformed massless free scalar field theory relates to the Nambu-Goto action in static gauge. The associated non-local property of $T\bar{T}$-deformed theories have already been discovered while studying the effective theory of long relativistic strings \cite{Dubovsky:2012wk, Dubovsky:2013ira}. Furthermore, the relation between the $T\bar{T}$-deformation theories and string theory have been intensively investigated in \cite{Frolov:2019nrr,Frolov:2019xzi,Sfondrini:2019smd,Callebaut:2019omt,Apolo:2019zai,Esper:2021hfq}.

Thanks to celestial holography, any quantum scattering amplitude of
massless particles in four-dimensional (4D) asymptotically Minkowskian spacetime can be rewritten as a correlation
function on the celestial sphere at null infinity \cite{Pasterski:2016qvg,Pasterski:2017kqt,Pasterski:2017ylz}, namely, celestial CFTs.  We argue that a 2D $T\bar{T}$ deformation of celestial CFTs can be applied to a 4D gravitational theory, which could shed light on the construction of a UV complete theory for General Relativity. We demonstrate our proposal by deriving a $T\bar{T}$ deformation soft graviton theorem in the context of celestial CFTs. Soft theorems and asymptotic symmetries are mathematically equivalent in many theories with massless particles, revealing the symmetry origin of universal factorization properties of scattering amplitudes in the soft limit \cite{Strominger:2013lka,Strominger:2013jfa,He:2014laa,Kapec:2014opa,He:2014cra,Campiglia:2014yka,He:2015zea,Campiglia:2016jdj,Campiglia:2016hvg,Conde:2016csj,Campiglia:2016efb,Conde:2016rom,Strominger:2017zoo,Campiglia:2017dpg,Mao:2017tey,Hamada:2018vrw,Anupam:2018vyu,Distler:2018rwu,Godazgar:2019dkh,Campiglia:2021bap,Liu:2021dyq,Miller:2022fvc}. In particular, the subleading soft graviton theorem\cite{Cachazo:2014fwa} implies that the tree-level S-matrix for quantum gravity in four-dimensional Minkowski space has Virasoro symmetry \cite{Kapec:2014opa}. Moreover, a 2D stress tensor was constructed from the subleading soft graviton theorem \cite{Kapec:2016jld}. It provokes the writing of scattering amplitudes in a basis \cite{Pasterski:2016qvg, Pasterski:2017kqt} manifesting the conformal symmetries. Then 4D tree-level scattering amplitudes are mapped to 2D correlators of CFTs on the celestial sphere \cite{Pasterski:2017ylz}, see also \cite{Raclariu:2021zjz,Pasterski:2021rjz,Pasterski:2021raf} for recent reviews and references therein. This connection allows one to deform a 4D theory with a 2D operator.

We start from the 2D charge defined by the stress tensor induced by the subleading soft graviton theorem \cite{Kapec:2016jld}. By introducing a soft graviton propagator, we can obtain the shadow of the subleading soft factor from the 2D charge. Then the subleading soft factor can be recovered by an inverse shadow transformation. The 2D stress tensor can be deformed by the $T\bar{T}$ operator in the standard way, which is given in a perturbative expansion of the deformation parameter $\lambda$. Accordingly, the deformed 2D charge leads to the shadow of the deformed soft factor. We perform the inverse shadow transformation and give the explicit form of the soft factor up to $\lambda^2$ order. The deformed soft theorem should be considered a universal factorization property of UV-complete quantum gravity. If confirmed by the ordinary amplitude calculation in momentum space, it provides remarkable evidence for the ongoing celestial holography program \cite{Raclariu:2021zjz, Pasterski:2021rjz, Pasterski:2021raf, string}.

\section{Soft theorem in asymptotic flat spacetime}
The undeformed theory lives on the asymptotic flat spacetime (AFS) background with retarded Bondi coordinates $(u,r,z,\bar{z})$. The AFS metric can be expanded near future null infinity $\mathcal{I}^+$ ($r\rightarrow\infty$)
\begin{equation}
    \begin{split} \label{eq:AFS4}
        ds^2= & - \frac{\mathring{R}}{2} du^2-2dudr+2r^2\gamma_{z\bar{z}}dzd\bar{z}\\
        &+\frac{2M}{r}du^2+rC_{zz}dz^2+rC_{\bar{z}\bar{z}}d\bar{z}^2\\
        & +D^zC_{zz} dudz + D^{\bar{z}} C_{\bar{z} \bar{z}} dud\bar{z}+\cdots,
    \end{split}
\end{equation}
where the retarded time $u=t-r$ is the coordinate of the null vector on $\mathcal{I}^+$. $D_z$ and $\mathring{R}$ are the covariant derivatives and Ricci scalar of the transverse metric $\gamma_{z \bar{z}}$ respectively. We choose $\gamma_{z\bar{z}}=\frac{2}{(1+z \bar{z})^2}$ for celestial sphere. Correspondingly, $\mathring{R}=2$. The asymptotic shear $C_{AB}$ and the Bondi mass aspect $M$ are independent of $r$. The Bondi news tensor is defined as follows
\begin{equation}
    N_{AB} = \p_u C_{AB}, \quad A,B = z\ \text{or} \ \bar{z}.
\end{equation}
One can use $\kappa h_{\mu \nu}\ (\kappa=\sqrt{32\pi G_N})$ to denote the perturbative part of AFS metric, which can be expanded as outgoing graviton modes
\be \label{eq:Mode expan}
    h^{out}_{\mu \nu} (x)= \sum_{\alpha=\pm} \int \frac{d^3q}{(2\pi)^3} \frac{1}{2\omega} \Big[\bar{\varepsilon}_{\mu\nu}^{\alpha} a_{\alpha}^{out}(\vec{q}) e^{i q \cdot x}\\
    +\varepsilon_{\mu\nu}^{\alpha}{a_{\alpha}^{out}(\vec{q})}^{\dagger} e^{-iq\cdot x}\Big],
\ee
where we have adopted natural units $8\pi G_N=1$, $q=(\omega, \vec{q})$ is the 4-momentum of graviton. The polarization tensor $\varepsilon_{\mu\nu}^{\pm }$ can be factorized as 2 polarization vectors $\varepsilon_{\mu\nu}^{\pm } =\varepsilon_{\mu}^{\pm} \varepsilon_{\nu}^{\pm}$. The momenta and the polarization vectors can be parametrized as
\begin{equation} \label{parametrization}
\begin{split}
&q^{\mu}(\omega,z,\bar{z})=\omega \bigg(1,\frac{z+\bar{z}}{1+z\bar{z}},\frac{-i(z-\bar{z})}{1+z\bar{z}},\frac{1-z\bar{z}}{1+z\bar{z}}\bigg),\\
&\varepsilon^{+}_{\mu}(q)=\frac{1}{\sqrt{2}}(-\bar{z},1,-i,-\bar{z}),\quad \varepsilon^-_{\mu}(q) =\bar{\varepsilon}^+_{\mu}(q).
\end{split}
\end{equation}
The canonical quantification gives
\begin{equation}
    \left[ a_{\alpha}^{out}(\vec{q}),{a_{\beta}^{out}(\vec{q'})}^{\dagger}\right]=2\omega_q \delta_{\alpha \beta} (2\pi)^3\delta^{(3)} (\vec{q}-\vec{q}').
\end{equation}
Comparing \eqref{eq:Mode expan} with \eqref{eq:AFS4}, one obtains the mode expansion for the shear and news tensors as
\begin{align}
    &C_{\bar{z}\bar{z}}=-\frac{i\hat{\varepsilon}_{\bar{z}\bar{z}}}{4\pi^2}\int^{\infty}_0 \td\omega \Big[a_-^{out}(q)e^{-i\omega u}-{a_+^{out}(q)}^{\dagger}e^{i\omega u}\Big],\notag\\
    &N_{\bar{z}\bar{z}}=-\frac{\hat{\varepsilon}_{\bar{z}\bar{z}}}{4\pi^2} \int^{\infty}_0\omega \td\omega \Big[a_-^{out}(q) e^{-i\omega u}+{a_+^{out}(q)}^{\dagger}e^{i\omega u}\Big],\notag\\
    & \hat{\varepsilon}_{\bar{z}\bar{z}}= \frac{2}{(1+z \bar{z})^2} = \gamma_{z\bar{z}},\nn
\end{align}
where we used the stationary-phase approximation \cite{Strominger:2017zoo}. Hence, the $n$-th moment of the news tensor can be written in the mode expansion as
\begin{equation}\label{moment}
    \begin{split}
        &N_{\bar{z}\bar{z}}^{(n)}=\int_{-\infty}^{\infty}du u^n N_{\bar{z}\bar{z}}\\
        &=\frac{(-i)^n}{2}\lim_{\omega\rightarrow 0}\partial_{\omega}^n\int^{\infty}_{-\infty} du \left(e^{i \omega u}+(-1)^n e^{-i \omega u}\right) N_{\bar{z}\bar{z}}\\
        &=-\frac{(-i)^n\hat{\varepsilon}^+_{\bar{z}\bar{z}}}{4\pi}\lim_{\omega\rightarrow 0}\p_{\omega}^n\left\{\omega \left[a_-^{out}(q)+(-1)^n{a_+^{out}(q)}^{\dagger}\right]\right\}
    \end{split}
\end{equation}

In the Heisenberg picture, the $n$-point tree level scattering problem in AFS can be regarded as that the asymptotic states $\ket{in}=\ket{q_1,s_1;\cdots ; q_m,s_m}$ defined on $\mathcal{I}^-$ and $\ket{out}=\ket{q_{m+1},s_{m+1};\cdots ; q_n,s_n}$ defined on $\mathcal{I}^+$ are fixed and $\mathcal{S}$-matrix depends on the time evolution. We use $q_k,s_k$ to denote the 4-momentum and helicity of $k$-th massless hard particle with finite energy $\omega_k$. The expression of $q_k,\varepsilon_{\mu}^{\pm} (q_k)$ can be similarly parametrized as \eqref{parametrization}. The $n$-point amplitude of massless hard particles is defined as follows
\begin{align}
    \mathcal{A}_n= \bra{out} \mathcal{S} \ket{in} =\<\mathcal{O}_1\cdots \mathcal{O}_n\>,
\end{align}
where $\mathcal{O}_k$ is annihilation or creation operator of $k$-th hard particle \cite{Kapec:2022axw},
\begin{equation*}
    \mathcal{O}_k(\omega_k,z_k,\bar{z}_k) = a_k^{out} (q_k) \theta(\omega_k) + a_k^{in \dagger} (-q_k) \theta(-\omega_k).
\end{equation*}
By introducing the following Mellin transform,
\begin{equation}
    \mathcal{O}_{\Delta_k,s_k} (z_k,\bar{z}_k) = \int_0^{\infty} d \omega_k \omega_k^{\Delta_k-1} \mathcal{O}_k(\omega_k,z_k,\bar{z}_k),
\end{equation}
one can connect the operators between 2D and 4D. Therefore, the 4D amplitude is equivalent to the 2D correlation function by implementing Mellin transform for all hard particles
\begin{equation}
    \<X_n\> = \prod^n_{k=1} \left(\int_0^{\infty} d\omega_k \omega_k^{\Delta_k-1} \right) \delta^{(4)}\left(\sum_{k=1}^{n} \epsilon_k q_k\right) \mathcal{A}_n,
\end{equation}
where $X_n = \prod^n_{k = 1} \mathcal{O}_{\Delta_k,s_k} (z_k,\bar{z}_k)$, and $\epsilon_k=1,-1$ for particles in $\ket{out}$ and $\ket{in}$ state respectively. See more details of Mellin transform in \cite{Pasterski:2017ylz}.

In terms of the soft theorem, an amplitude containing $n$ hard particles and a soft graviton with energy $\omega \rightarrow 0$ can be expanded by the power of soft energy $\omega$
\be
\begin{split}
    \mathcal{A}_{n+1}^{\pm}(q) \big|_{\omega \rightarrow 0} &= \lim_{\omega \rightarrow 0} \bra{out;q,\pm2} \mathcal{S} \ket{in} \\
    &= \lim_{\omega \rightarrow 0}\bra{out}a_{\pm}^{out}(q)\mathcal{S}\ket{in} = \sum^{\infty}_{n=0} S^{(n)\pm} \omega^{n-1} \mathcal{A}_n,
\end{split}
\ee
Since the $n$-th news tensor can be expressed by the generator of $n$-th order soft factor \eqref{moment}, one can read off the results of the insertion of the $n$-th moment into the amplitude as \begin{equation}
    \bra{out} N^{z(n)}_{\ \ \bar{z}}\mathcal{S} \ket{in} = -\frac{(-i)^n n!}{4\pi} S^{(n)-}\bra{out}\mathcal{S}\ket{in}.
\end{equation}
The soft factors are universal for the first three orders \cite{Cachazo:2014fwa}. In particular,  the subleading soft graviton factors can be written in the position space as \cite{Kapec:2016jld}
\begin{align}
S^{(1)+}_{(\vec{z},\vec{z}_k)}&=\sum^{n}_{k=1}\frac{(\bar{z}-\bar{z}_k)^2}{z-z_k}\bigg[\frac{2\hat{\bar{h}}_k}{\bar{z}-\bar{z}_k}-\Gamma^{\bar{z}_k}_{\bar{z}_k\bar{z}_k}\hat{\bar{h}}_k -\p_{\bar{z}_k}-s_k\Omega_{\bar{z}_k}\bigg],\notag\\
S^{(1)-}_{(\vec{z},\vec{z}_k)}&=\sum^{n}_{k=1}\frac{(z-z_k)^2}{\bar{z}-\bar{z}_k}\bigg[\frac{2\hat{h}_k}{z-z_k}  -\Gamma^{z_k}_{z_k z_k}\hat{h}_k -\p_{z_k}+s_k\Omega_{z_k}\bigg],\notag\\
\hat{h}_k&=\frac{1}{2}(s_k-\omega_k\p_{\omega_k}),\quad\hat{\bar{h}}_k=\frac{1}{2}(-s_k-\omega_k\p_{\omega_k})\label{S-}
\end{align}
where $\vec{z}=(z,\bar{z})$ and $\vec{z}_k=(z_k,\bz_k)$ are the locations of the soft graviton and the hard particles respectively, $\Gamma^z_{zz}$ is the Levi-Civita connection of the celestial sphere metric $\gamma_{z\bar{z}}$ and $\Omega_z=\frac{\Gamma^z_{zz}}{2}$ is the spin connection.

\section{From 4D superrotation charge to 2D Virasoro charge}

In 4D AFS, the gravitational scattering has Bondi-Metzner-Sachs (BMS) invariance \cite{Strominger:2013jfa}, which reveals the symmetry origin of the soft graviton theorem. The BMS symmetry consists of supertranslations and superrotations related to the leading \cite{He:2014laa} and subleading \cite{Kapec:2014opa, Campiglia:2014yka} soft graviton theorem, respectively. The superrotation charge includes two parts, namely the soft part and the hard part \cite{Kapec:2014opa, Campiglia:2014yka, Raclariu:2021zjz, Pasterski:2021rjz},
\be
\mathcal{Q}=\mathcal{Q}_S+\mathcal{Q}_H,
\ee
which are given by
\be\label{Qs}
\begin{split}
&{\cal Q}_H=-2i\int_{{\cal I}^+} \gamma_{z\bz} \td^2 z \td u \left(Y^z T_{uz}^{(4)} + u D_z Y^z T_{uu}^{(4)}\right),\\
&{\cal Q}_S= i \int_{{\cal S}^2} \td^2 z Y^z D_z^3 N^{z(1)}_{\ \ \bz} = i \int \td^2 z Y^z\p_z^3 N^{z(1)}_{\ \ \bz},
\end{split}
\ee
for the holomorphic case, where $T_{\mu \nu}^{(4)}$ is the 4D total stress tensor \cite{Strominger:2017zoo}, $Y^A$ is the superrotation parameter. The Ward identity of the superrotation charge yields the insertion of soft charge $\mathcal{Q}_S$ as \cite{Kapec:2014opa}
\be \label{eq:<QY>}
    \bra{out} \mathcal{Q}_S \mathcal{S} \ket{in} = \sum_{k=1}^n \Big(Y^{z_k}(\p_{z_k}-s_k\Omega_{z_k})+D_{z_k} Y^{z_k} \hat{h}_k\Big)\\
    \times \bra{out} \mathcal{S} \ket{in}.
\ee
Applying Mellin transform for hard particles in (\ref{eq:<QY>}) yield
\begin{equation} \label{eq:2D<calQY>}
    \< Q_Y X_n\>=\sum_{k=1}^n\Big[Y^{z_k}(\p_{z_k}-s_k\Omega_{z_k})+D_{z_k}Y^{z_k}h_k\Big] \< X_n \>.
\end{equation}
where $(h_k,\bar{h}_k)=(\frac{\Delta_k+s_k}{2},\frac{\Delta_k-s_k}{2})$ are the conformal weights of $k$-th hard particle on the celestial sphere. The subscript $Y$ in the charge indicates that it corresponds to a 2D charge operator. Remarkably, eq.\eqref{eq:2D<calQY>} recovers the Ward identity of 2D Virasoro charge constructed in \cite{Kapec:2016jld} from 2D stress tensor,
\begin{equation} \label{QY}
    Q_{Y}=\frac{1}{2\pi i}\oint_{\cal C} \td z T_{z z} Y^{z},
\end{equation}
where the integral contour $\mathcal{C}$ separates the locations of all hard particles $\vec{z}_k$ and soft particles $\vec{z}$. The insertion of the 2D stress tensor in \eqref{QY} into the correlator yields \cite{Kapec:2016jld}
\be \label{eq:Ward}
\< T_{zz} X_n \> = \sum^n_{k=1} \bigg[\frac{h_k}{(z-z_k)^2} +\frac{\Gamma^{z_k}_{z_kz_k}}{z-z_k}h_k\\
+\frac{1}{z-z_k}(\p_{z_k}-s_k\Omega_{z_k})\bigg] \< X_n \>,
\ee
which is precisely the conformal Ward Identity of stress tensor on celestial sphere \cite{Eguchi:1986sb}. While the OPEs of the stress tensor are derived by inserting 2 components of the stress tensor into the amplitude and implementing Mellin transform \cite{Fotopoulos:2019vac},
\begin{equation}\begin{split}
&T_{zz}T_{z'z'}\sim\frac{2T_{z'z'}}{(z-z')^2}+\frac{\p_{z'}T_{z'z'}}{z-z'}+\text{regular},\\ &T_{zz}T_{\bar{z}'\bar{z}'}\sim \text{regular},
\end{split}\end{equation}
which indicates that the central charge of the corresponding CFT on the celestial sphere is vanishing and the stress tensor is traceless.

Following the above procedures, one can easily recover the correspondence between the 4D and 2D charges for the anti-holomorphic part.

\section{From 2D Virasoro charge to subleading soft graviton theorem}

Superrotations reveal the symmetry origin of the subleading soft graviton theorem \cite{Kapec:2014opa, Campiglia:2014yka, Campiglia:2015yka}. Hence, the 2D Virasoro charge is related to the subleading soft graviton theorem in the context of celestial holography \cite{Raclariu:2021zjz, Pasterski:2021rjz, Pasterski:2021raf, string}. It is shown that soft theorems can be directly derived in the 2D conformal basis \cite{Adamo:2019ipt, Puhm:2019zbl, Guevara:2019ypd}. Here, we propose a direct way to reveal the subleading conformally soft graviton theorem \cite{Adamo:2019ipt, Puhm:2019zbl, Guevara:2019ypd} from 2D Virasoro charge. The 2D stress tensor with dimension $\Delta=2$ is the shadow transformation of the subleading soft-graviton operator \cite{Kapec:2017gsg} with dimension $\Delta=0$ \cite{Adamo:2019ipt, Puhm:2019zbl, Guevara:2019ypd}. One can refer to the shadow transformation in Appendix \ref{shadow} and also in \cite{Dolan:2011dv, Pasterski:2017kqt, Donnay:2018neh, Haehl:2019eae}. Thanks to this shadow relation, one can verify that the 2D Virasoro charge associated with a particular choice \cite{Strominger:2017zoo} of the superrotation parameter $Y^z=\frac{1}{w-z}$ is the shadow of the subleading soft-graviton operator. Further, one can apply the same choice for superrotation charge to recover the subleading soft graviton theorem \cite{Kapec:2014opa, Strominger:2017zoo}. We would refer to the particular choice $Y^z$ \cite{Strominger:2017zoo} as a soft graviton propagator. In such a way, the 2D Virasoro charge reveals the symmetry origin of the conformally soft graviton theorem \cite{Adamo:2019ipt, Puhm:2019zbl, Guevara:2019ypd} in the 2D context. It can be justified by inserting the Mellin transform of the 2D charge associated with the soft graviton propagator into a 4D amplitude,
\begin{equation}
    \bra{out}Q_{Y}\mathcal{S}\ket{in}= \widetilde{S}^{(1)-}_{ (\vec{w})} \bra{out}\mathcal{S}\ket{in},
\end{equation}
where
\begin{equation} \label{eq:widetildeS}
\begin{split}
    &\widetilde{S}^{(1)-}_{ (\vec{w})}=\sum^n_{k=1}\bigg[\frac{\hat{h}_k}{(w-z_k)^2}+\frac{\Gamma^{z_k}_{z_kz_k} \hat{h}_k}{w-z_k}+\frac{\p_{z_k}-s_k\Omega_{z_k}}{w-z_k}\bigg],\\
    &\widetilde{S}^{(1)+}_{ (\vec{w})}=\sum^n_{k=1}\bigg[\frac{\hat{\bar{h}}_k}{(\bar{w}-\bar{z}_k)^2}+\frac{\Gamma^{\bar{z}_k}_{\bar{z}_k\bar{z}_k} \hat{\bar{h}}_k}{\bar{w}-\bar{z}_k}+\frac{\p_{\bar{z}_k}+s_k\Omega_{\bar{z}_k}}{\bar{w}-\bar{z}_k}\bigg].
\end{split}
\end{equation}
The factor $\widetilde{S}^{(1)-}_{ (\vec{w},\vec{z}_k)}$ is related to soft factor $S^{(1)-}_{(\vec{z},\vec{z}_k)}$ in \eqref{S-}
by the shadow transformation as
\begin{equation} \label{eq:ShadowofS}
    \begin{split}
        \widetilde{S}^{(1)-}_{(\vec{w})}&=\frac{3!}{4\pi}\int \td^2 z \frac{1}{(w-z)^4} {S}^{(1)-}_{(\vec{z})},\\
        S^{(1)-}_{(\vec{z})}&=\frac{1}{2\pi}\int \td^2w\frac{(z-w)^2}{(\bar{z}-\bar{w})^2}\widetilde{S}^{(1)-}_{(\vec{w})}.
    \end{split}
\end{equation}

\section{2D charges from T\={T} deformed stress tensor}

The $T\bar{T}$ flow effect on action is
\begin{equation}\label{flow}
    \frac{\p S^{[\lambda]} }{\p \lambda} = -\int \sqrt{ \gamma^{[\lambda]} } d^2x O^{[\lambda]}_{T \overline{T}},\quad O^{[\lambda]}_{T \overline{T}}=\frac{1}{2} (T^{AB}_{[\lambda]} T^{[\lambda]}_{AB} -T^2_{[\lambda]}),
\end{equation}
where $\lambda$ is the coupling constant of $T \bar{T}$ deformation and $T$ denotes the trace of stress tensor. The superscript $[0]$ denotes quantities before $T\bar{T}$ deformation, while $[\lambda]$ denotes the deformed quantities. The flow equation can be exactly solved with variational principle \cite{Guica:2019nzm}
\begin{equation}
    \begin{split}
        &\hat{T}_{AB}^{[\lambda]}  = T_{AB}^{[\lambda]}-\gamma_{AB}^{[\lambda]} T^{[\lambda]} =\hat{T}_{AB}^{[0]} -\hat{T}_{AC}^{[0]}\hat{T}_{BD}^{[0]}\gamma^{CD}_{[0]}\lambda, \\
        &\gamma_{AB}^{[\lambda]}=\gamma_{AB}^{[0]}-2\hat{T}_{AB}^{[0]}\lambda +\hat{T}_{AC}^{[0]}\hat{T}_{BD}^{[0]}\gamma^{CD}_{[0]} \lambda^2,\\
        &T^{[\lambda]} =\frac{T^{[0]}-2O_{T\overline{T}}^{[0]} \lambda}{1+T^{[0]} \lambda-O_{T\overline{T}}^{[0]} \lambda^2}.
    \end{split}
\end{equation}
One can check that the deformed stress tensor is conserved, namely $\gamma^{AB}_{[\lambda]} D_{A}^{[\lambda]} T^{[\lambda]}_{BC}=0$, where $ D_{A}^{[\lambda]}$ is the covariant derivative with respect to the deformed metric $\gamma_{AB}^{[\lambda]}$. On the celestial sphere, the perturbative terms of components of the deformed stress tensor are
\begin{equation}\label{eq:T[l]zz,zzbar,zbarzbar}
    \begin{split}
        T^{[\lambda]}_{zz}&=T^{[0]}_{zz}\Big[1+4\sum_{n=1}^{\infty}(O_{T\overline{T}}^{[0]}\lambda^2)^n\Big],\\
        T^{[\lambda]}_{\bar{z}\bar{z}}&=T^{[0]}_{\bar{z}\bar{z}}\Big[1+4\sum_{n=1}^{\infty}(O_{T\overline{T}}^{[0]}\lambda^2)^n\Big],\\
        T^{[\lambda]}_{z\bar{z}}&=-\gamma^{[0]}_{z\bar{z}}\Big[3+4\sum_{n=1}^{\infty}(O_{T\overline{T}}^{[0]}\lambda^2)^n\Big]O_{T\overline{T}}^{[0]}\lambda.
    \end{split}
\end{equation}
For the deformed stress tensor, the corresponding 2D charge is
\begin{equation}
    Q^{[\lambda]}=\frac{1}{2\pi i}\oint_{\cal C} \td x^AT^{[\lambda]}_{AB} Y^B= Q^{[\lambda]}_{Y}+Q^{[\lambda]}_{\bar{Y}},
\end{equation}
where
\begin{equation} \label{holomorphic}
    \begin{split}
        &Q^{[\lambda]}_{Y}=\frac{1}{2\pi i}\oint_{\cal C} \td z T^{[\lambda]}_{zz}Y^z + \frac{1}{2\pi i}\oint_{\cal C} \td\bar{z} T^{[\lambda]}_{z\bar{z}} Y^z,\\
        &Q^{[\lambda]}_{\bar{Y}}=\frac{1}{2\pi i}\oint_{\cal C} \td \bz T^{[\lambda]}_{\bz \bz}Y^{\bz} + \frac{1}{2\pi i} \oint_{\cal C} \td z T^{[\lambda]}_{z\bar{z}}Y^{\bz} .
    \end{split}
\end{equation}
Inserting the deformed stress tensor \eqref{eq:T[l]zz,zzbar,zbarzbar}, we obtain the  minus helicity charge \eqref{holomorphic} in series expansion of $\lambda$ as
\begin{multline}\label{deformedcharge}
Q^{[\lambda]}_{Y}=\oint\frac{\td z}{2\pi i} Y^z T^{[0]}_{zz}-3\lambda\oint\frac{\td\bar{z}}{2\pi i}Y^z\gamma^{z\bar{z}[0]}T^{[0]}_{z z}T^{[0]}_{\bar{z}\bar{z}}\\
-4\sum_{s=1}^{\infty}\lambda^{2s+1}\oint\frac{\td\bar{z}}{2\pi i}Y^z(T^{[0]}_{zz})^{s+1}(T^{[0]}_{\bar{z}\bar{z}})^{s+1}(\gamma^{z\bar{z}}_{[0]})^{2s+1}\\
+4\sum_{s=1}^{\infty}\lambda^{2s}\oint\frac{\td z}{2\pi i}Y^z (T^{[0]}_{zz})^{s+1} (T^{[0]}_{\bar{z}\bar{z}})^{s} (\gamma^{z\bar{z}}_{[0]})^{2s}.
\end{multline}
To close this section, two remarks about the deformed charges are as follows.
Firstly, since the deformed stress tensor has three independent components, one cannot directly apply the connection \cite{Kapec:2016jld} between the 2D traceless stress tensor and the subleading soft graviton theorem to our case. Alternatively, we construct the deformed charge by contracting the deformed stress tensor with the superrotation (conformal killing) vectors. Secondly, since the contour integration associated with the $\lambda$ odd order terms in \eqref{deformedcharge} is irrelevant to minus helicity soft graviton propagator $Y^z$, the integration contour can not attach the soft graviton propagator to any hard particles. It corresponds to disconnected correlators, namely a soft graviton propagator plus the correlation function of hard particles.

\section{T\={T} deformed soft theorem}

The deformed charges play essential roles in obtaining the deformed soft graviton theorem. In particular, one can insert the charge \eqref{deformedcharge} into the amplitudes to get the shadow of a deformed subleading minus helicity soft graviton theorem. In the Heisenberg picture, the asymptotic states $\bra{out}$ and $\ket{in}$ are the same as the ones in un-deformed theory, the information of deformation is hidden in $\mathcal{S}$-matrix,
\begin{equation}
    \mathcal{A}_n^{[\lambda]} = \bra{out} \mathcal{S}^{[\lambda]} \ket{in}.
\end{equation}
As discussed previously, after the insertion of $Q_{Y}^{[\lambda]}$ into the amplitudes, together with $Y^z=\frac{1}{w-z}$, the shadow of the soft factor will be obtained by shrinking the integral contour away from locations of hard particles $z_k$,
\be\begin{split}
&\widetilde{S}^{[\lambda](1)-}_{(\vec{w},\vec{z}_k)} \bra{out} \mathcal{S}^{[\lambda]} \ket{in}=\bra{out}Q_{Y}^{[\lambda]}\mathcal{S}^{[\lambda]}\ket{in}
\end{split}\ee
which is equivalent to the insertion of $T^{[\lambda]}_{ww}$ and subtracting all extra $\delta$ functions because the integral contour does not pass through the poles of the delta function. The shadow of the deformed soft factor is given by
\be\label{eq:tildeS(1)-}
    \widetilde{S}^{(1)-}_{[\lambda](\vec{w},\vec{z}_k)}=\widetilde{S}^{(1)-}_{ (\vec{w},\vec{z}_k)}
    +4\sum_{s=1}^{\infty}\lambda^{2s}(\gamma^{z\bar{z}}_{[0]})^{2s}\left[\widetilde{S}^{(1)-}_{ (\vec{w},\vec{z}_k)}\right]^{s+1}\left[\widetilde{S}^{(1)+}_{ (\vec{w},\vec{z}_k)}\right]^s,
\ee
where $\widetilde{S}^{(1)\pm}_{ (\vec{w},\vec{z}_k)}$ is the shadow of undeformed subleading soft factor \eqref{eq:widetildeS}. The $T\bar{T}$ deformation doesn't change the helicity of $T_{zz}$ and $T_{\bz\bz}$, and they follow the same shadow formula \eqref{eq:ShadowofS}. The explicit formula of the soft factor is
\begin{multline} \label{deformedsoftfactor}
    S^{(1)-}_{[\lambda]}(\vec{z},\vec{z}_k)= S^{(1)-}_{ (\vec{z},\vec{z}_k)}+\frac{2}{\pi}\sum_{s=1}^{\infty}\lambda^{2s}\int \td^2 w\frac{(z-w)^2}{(\bar{z}-\bar{w})^2}\\
    \times(\gamma^{z\bar{z}}_{[0]})^{2s} \left[\widetilde{S}^{(1)-}_{ (\vec{w},\vec{z}_k)}\right]^{s+1}\left[\widetilde{S}^{(1)+}_{ (\vec{w},\vec{z}_k)}\right]^s,
\end{multline}
where $S^{(1)-}_{ (\vec{z},\vec{z}_k)}$ is the undeformed subleading soft factor \eqref{S-}. The leading order of $\lambda$ expansion restores the undeformed soft factor. The explicit forms at $\lambda^2$ order and the strategy of computing surface integral on the celestial sphere in the complex stereographic coordinates are presented in Appendix \ref{detail1}. Finally, the soft factor can be translated into momentum space with the relation between the celestial sphere coordinates and null momenta and polarization vectors in \eqref{parametrization}. Similarly, the plus helicity soft graviton theorem can be obtained from the antiholomorphic charge $Q_{\bar{Y}}$ defined in \eqref{holomorphic}.

\section{Scheme for deriving T\={T} deformed amplitudes}

In this paper, we propose to employ 2D $T\bar{T}$ deformation to explore the UV property of 4D scattering amplitudes in the context of celestial holography. As the first example, we illustrate the proposal by deriving a $T\bar{T}$ deformed subleading soft graviton theorem.
Because the connection of 4D amplitudes and 2D correlators is implemented by the Mellin transform, which is a sum of all energies that mixes the infrared and ultraviolet regimes, one can not directly perform the Mellin transform to a soft theorem. At the same time, the soft theorem has its clear symmetry origin and can be recast as the Ward identity of asymptotic symmetry. Then the soft theorem in celestial holography can be regarded as the correspondence of two Ward identities in 2D and 4D, respectively.

The final goal of our proposal is to find the $T\bar{T}$ deformed full amplitude and to study its UV property. Our derivation of $T\bar{T}$ deformed soft theorem can already fix the deformed full amplitudes for some particular theories whose amplitudes are determined by their soft theorems \cite{Cachazo:2016njl,Rodina:2018pcb}, because the deformed soft factor only consists of undeformed soft factors. For a generic theory, one should first study the deformed correlators and perform the inverse Mellin transformation to obtain the deformed amplitude. Since irrelevant deformation flows to higher energy scales, the $T\bar{T}$ deformation should have certain relations to the loop correction from the amplitude side. Presumably, we will show some overlapping sectors between the $T\bar{T}$ deformed correlators and loop corrections in 4D amplitudes with proper assumptions.
In particular, the first-order deformed CFT correlators can be expressed by the undeformed CFT correlators as \cite{Cardy:2019qao}
\begin{equation}\label{firstcorrection}
 \<X_n\>^{(1)}_{[\lambda]}|_{\text{div}} = 2 \lambda \sum_{i \neq j} \left( \log \frac{\epsilon^2}{|Z_{ij}|^2} \right) \p_{\bar{Z}_j} \p_{Z_i}  \<X_n\>_{[0]},
\end{equation}
where $\epsilon$ is a regularization parameter from the point-splitting, and $Z_{ij} = Z_i - Z_j$. Here the CFT coordinates $(Z,\bar{Z})$ are the same as $(z,\bar{z})$ defined in previous sections if we use the natural unit, namely, setting the dimensionful constant to 1, see more discussion in Appendix \ref{detail2}.
On the 4D gravity side, the one-loop-order form of infrared singularities \cite{Dunbar:1995ed, Naculich:2011ry, Akhoury:2011kq} in gravity amplitude is following
\be\label{eq:amplitude}
\mathcal{A}^{\text{1-loop}}_n|_{\text{leading div}}= \frac{1}{\Lambda} \sigma_n \mathcal{A}^{\text{tree}}_n,
\ee
where
\be
\sigma_n=-c_\Gamma \sum_{i=0}^{n-1}\sum_{j=i+1}^{n} (p_i+p_j)^2 \log \left(\frac{\mu^2}{-(p_i+p_j)^2}\right),
\ee
$\Lambda$ is the dimensional regularization parameter, $\mu^2$ is the usual dimensional regularization scale, and $p_i$ is the momentum of the $i$-th particle.

To manifest the connection of $T\bar{T}$ deformation and loop correction, we perform Mellin transforms on the amplitude relation \eqref{eq:amplitude}, which yields
\be \label{eq:2Dloop}
    \<X_n\>^{\text{leading-div}}_{(1)} =   - \frac{G_N}{\pi \epsilon} \sum_{i, j = 1}^n \log \left( \frac{|Z_{ij}|^2}{\Lambda_{ij}^2} \right) ( 2\p_{Z_i} \p_{\Bar{Z}_j} - T_i T_j ) \<X_n\>^{[0]},
\ee
where $ \<X_n\>^{[0]}$ is related to the tree level amplitude in the form of Mellin transform and the action of the operator $T_k$ is
\begin{equation}
    T_k \mathcal{O}_{\Delta_k, s_k} (\Vec{z}_k) = - i \mathcal{O}_{\Delta_k+1, s_k} (\Vec{z}_k).
\end{equation}
In the course of the Mellin transform, we have applied the recently discovered connection between the conformal Carrollian field theory and celestial CFT \cite{Donnay:2022aba, Donnay:2022wvx}. The details of the transform are given in Appendix \ref{detail2}. If one identifies the Newton constant $G_N$ and the deformation parameter $\lambda$ as
\begin{equation}
    \lambda = - \frac{2 G_N}{\pi^2 \epsilon},
\end{equation}
the 2D correlator relation \eqref{firstcorrection} is part of the Mellin transform of the amplitude relation \eqref{eq:amplitude}. In principle, one should not expect that the $T\bar{T}$ deformation is completely equivalent to the loop correction in the context of celestial holography. Because the $T\bar{T}$ deformation determines a UV complete theory while the gravity theory with loop corrections is not. Nevertheless, those computations are indicating that the $T\bar{T}$ deformation and the loop correction are on the same footing to approach the high energy scale effect. We leave their concrete relation for future investigation. And another natural question is how one can make a soft theorem compatible with the Mellin transform and study the loop-corrected soft theorem \cite{Bern:2014oka, He:2014bga} (see also \cite{He:2017fsb, Pasterski:2022djr, Donnay:2022hkf} for the corrected stress tensor from the soft theorem) in the context of the 2D $T\bar{T}$ deformation. Furthermore, it is interesting to study the deformed charge algebra based on our current construction of the deformed charges.

\section*{Acknowledgments}
The authors thank Yi-hong Gao, Miao He, Yi Li, Zhengwen Liu, Sabrina Pasterski, Romain Ruzziconi, Yuan Sun, and Yu-Xuan Zhang for valuable discussions and comments. We also thank two anonymous reviewers for drawing our attention to loop corrections of the amplitudes. This work is partly supported by the National Natural Science Foundation of China under Grant No.~12075101, No.~12235016, No.~11905156 and No.~11935009. S.H. would appreciate the financial support from Jilin University and Max Planck Partner Group.

\appendix

\section{Shadow transformation}
\label{shadow}

\subsection{Definition}

The shadow of a $d$-dimensional scalar operator $\mathcal{O}_{\Delta}$ with conformal dimension $\Delta$ is defined as \cite{Dolan:2011dv,Pasterski:2017kqt,Donnay:2018neh,Haehl:2019eae}
\begin{equation}
    \widetilde{\mathcal{O}}(\vec{x})=\frac{\Gamma(\Delta)}{\pi^{\frac{d}{2}}\Gamma(\Delta-\frac{d}{2})}\int d^d\vec{x}'\frac{\mathcal{O}_{\Delta}(\vec{x}')}{|\vec{x}-\vec{x}'|^{2(d-\Delta)}}.
\end{equation}

When restricted on the celestial sphere, one has
\begin{equation}\label{eq:2d scalar shadow}
    \widetilde{\mathcal{O}}(\vec{x})=\frac{\Gamma(\Delta)}{\pi\Gamma(\Delta-1)}\int d^2\vec{x}'\frac{\mathcal{O}_{\Delta}(\vec{x}')}{|\vec{x}-\vec{x}'|^{2(2-\Delta)}}.
\end{equation}
Following \cite{Donnay:2018neh}, the shadow of symmetric-traceless operator with spin $J$ can be defined as
\begin{equation}
    \widetilde{\mathcal{O}}_{a_1\cdots a_{|J|}}(\vec{x})=\frac{k_{\Delta,J}}{\pi}\int d^2\vec{x}'\frac{I_{a_1b_1}(\vec{x}-\vec{x}')\cdots I_{a_{|J|}b_{|J|}}(\vec{x}-\vec{x}')}{|\vec{x}-\vec{x}'|^{2(2-\Delta)}}\mathcal{O}^{b_1\cdots b_{|J|}}_{\Delta}(\vec{x}'),
\end{equation}
where
\begin{equation}
    I_{ab}(\vec{x})=\delta_{ab}-2\frac{x_ax_b}{|\vec{x}|^2}.
\end{equation}
Properly choosing the normalization factor $k_{\Delta,J}$ can ensure that  \cite{Dolan:2011dv}
\begin{equation}
    \widetilde{\widetilde{\mathcal{O}}}_{a_1\cdots a_{|J|}}(\vec{x})=\mathcal{O}_{\Delta;a_1\cdots a_{|J|}}(\vec{x}).
\end{equation}
For notational brevity, we will refer to the second shadow transformation that  transforms back to the original operator as the inverse shadow transformation.

\subsection{Shadow of the stress tensor}

The shadow of a 2D stress tensor is defined as
\begin{equation}
\widetilde{T}_{a_1a_2}(\vec{x})=\frac{ik_{2,2}}{2\pi}\int d^2z'\mathcal{I}_{a_1a_2},
\end{equation}
where
\begin{equation}\label{eq:Ia1a2}
\begin{split}
\mathcal{I}_{a_1a_2}(\vec{x},\vec{x}')=&I_{a_1b_1}(\Delta\vec{x})I_{a_2b_2}(\Delta\vec{x})T^{b_1b_2}(\vec{x}')\\
=&\bigg[\delta_{a_1b_1}-2\frac{\Delta x_{a_1}\Delta x_{b_1}}{(z-z')(\bar{z}-\bar{z}')}\bigg]\bigg[\delta_{a_2b_2}-2\frac{\Delta x_{a_2}\Delta x_{b_2}}{(z-z')(\bar{z}-\bar{z}')}\bigg]T^{b_1b_2}(\vec{x}')\\
=&T_{a_1a_2}(\vec{x}')-2\frac{\Delta x_{a_2}\Delta x^{b_2}T_{a_1b_2}(\vec{x}')+\Delta x_{a_1}\Delta x^{b_1}T_{b_1a_2}(\vec{x}')}{(z-z')(\bar{z}-\bar{z}')}\\
&\quad +4\frac{\Delta x_{a_1}\Delta x_{a_2}\Delta x^{b_1}\Delta x^{b_2}T_{b_1b_2}(\vec{x}')}{(z-z')^2(\bar{z}-\bar{z}')^2},
\end{split}
\end{equation}
and we define $\Delta x=x-x'$.

Setting $a_1=a_2=z$ yields
\begin{align*}
\mathcal{I}_{zz}=&T_{zz}(\vec{x}')-4\frac{\Delta x_{z}\Delta x^{b_2}T_{zb_2}(\vec{x}')}{(z-z')(\bar{z}-\bar{z}')}+4\frac{(\Delta x_{z})^2\Delta x^{b_1}\Delta x^{b_2}T_{b_1b_2}(\vec{x}')}{(z-z')^2(\bar{z}-\bar{z}')^2}\\
=&T_{zz}(\vec{x}')-2\frac{\Delta x^{z}T_{zz}(\vec{x}')+\Delta x^{\bar{z}}T_{z\bar{z}}(\vec{x}')}{z-z'}\\
&\quad +\frac{(\Delta x^{z})^2T_{zz}+(\Delta x^{\bar{z}})^2T_{\bar{z}\bar{z}}+2\Delta x^{z}\Delta x^{\bar{z}}T_{z\bar{z}}(\vec{x}')}{(z-z')^2}\\
=&\frac{(\bar{z}-\bar{z}')^2}{(z-z')^2}T_{\bar{z}\bar{z}}(\vec{x}')=\frac{(\bar{z}-\bar{z}')^2}{(z-z')^2}T_{\bar{z}\bar{z}}(z',\bar{z}').
\end{align*}
Similarly,
\begin{equation}
    \mathcal{I}_{\bar{z}\bar{z}}=\frac{(z-z')^2}{(\bar{z}-\bar{z}')^2}T_{zz}(z',\bar{z}').
\end{equation}
Hence
\begin{equation}
    \begin{split}
        &\widetilde{T}_{zz}=\frac{ik_{2,2}}{2\pi}\int d^2z'\mathcal{I}_{zz}=\frac{ik_{2,2}}{2\pi}\int d^2z'\frac{(\bar{z}-\bar{z}')^2}{(z-z')^2}T_{\bar{z}\bar{z}}(z',\bar{z}'),\\
        &\widetilde{T}_{\bar{z}\bar{z}}=\frac{ik_{2,2}}{2\pi}\int d^2z'\mathcal{I}_{\bar{z}\bar{z}}=\frac{ik_{2,2}}{2\pi}\int d^2z'\frac{(z-z')^2}{(\bar{z}-\bar{z}')^2}T_{zz}(z',\bar{z}'),
    \end{split}
\end{equation}
which recovers the result in \cite{Haehl:2019eae}.

On the celestial sphere, the shadow of the stress tensor is the “twice” shadow of the first moment of the Bondi news. If the normalization factor is chosen properly, the shadow of the stress tensor will yield the soft fact directly. We now verify this
\begin{equation}
\begin{split}
\widetilde{T}_{\bar{z}\bar{z}}=&\frac{ik_{2,2}}{2\pi}\int d^2z'\frac{(z-z')^2}{(\bar{z}-\bar{z}')^2}T_{zz}(z')\\
=&\frac{3!k_{2,2}}{2\pi}\int d^2z'd^2w\frac{(z-z')^2}{(\bar{z}-\bar{z}')^2}\frac{(\gamma^{z\bz}{\cal N}_{\bz\bz}^{(1)})(w,\bar{w})}{(z'-w)^4}\\
=&\frac{k_{2,2}}{2\pi}\int d^2z'd^2w\frac{(\gamma^{z\bz}{\cal N}_{\bz\bz}^{(1)})(w,\bar{w})}{z'-w}\partial_{z'}^{3}\frac{(z-z')^2}{(\bar{z}-\bar{z}')^2}\\
=&\frac{k_{2,2}}{2\pi}\int d^2z'd^2w\frac{(\gamma^{z\bz}{\cal N}_{\bz\bz}^{(1)})(w,\bar{w})}{z'-w}\partial_{\bar{z}'}\partial_{z'}^{3}\frac{(z-z')^2}{\bar{z}-\bar{z}'}\\
=&-2k_{2,2}\int d^2z'd^2w\frac{(\gamma^{z\bz}{\cal N}_{\bz\bz}^{(1)})(w,\bar{w})}{z'-w}\partial_{\bar{z}'}\delta^{(2)}(z-z')\\
=&2k_{2,2}\int d^2z'd^2w{(\gamma^{z\bz}\cal N}_{\bz\bz}^{(1)})(w,\bar{w})\delta^{(2)}(z-z')\partial_{\bar{z}'}\frac{1}{z'-w}\\
=&4\pi k_{2,2}\int d^2z'd^2w{(\gamma^{z\bz}\cal N}_{\bz\bz}^{(1)})(w,\bar{w})\delta^{(2)}(z-z')\delta^{(2)}(z'-w)\\
=&4\pi k_{2,2}\int d^2w(\gamma^{z\bz}{\cal N}_{\bz\bz}^{(1)})(w,\bar{w})\delta^{(2)}(z-w)\\
=&4\pi k_{2,2}\gamma^{z\bz}{\cal N}_{\bz\bz}^{(1)}(z,\bar{z}).
\end{split}
\end{equation}
Therefore,
\begin{equation}\label{eq:k2,2=-i}
k_{2,2}=-i\Leftrightarrow \widetilde{T}_{\bar{z}\bar{z}}=-4\pi i\gamma^{z\bz}{\cal N}_{\bz\bz}^{(1)}(z,\bar{z}),
\end{equation}
which will ensure that the insertion of the shadow of the stress tensor into the correlation function will yield the soft factor.

\subsection{Shadow of the deformed stress tensor}

After the $T\bar{T}$ deformation, the stress tensor is no longer traceless. We will split the deformed stress tensor into its trace part for which we denote as $\Theta$ and traceless part for which we denote as ${\cal G}_{AB}$. Then we can define the shadow of $\Theta$ and ${\cal G}_{AB}$ in different ways according to their conformal weights. Another effect from $T\bar{T}$ deformation is that the full stress tensor is not primary. We will follow the prescription in \cite{Haehl:2019eae} to simply extend the shadow operation to any traceless tensor.

\section{Surface integrals in the deformed soft factor}
\label{detail1}

The explicit form of the complete deformed soft factor can be expressed in terms of the following integral notation \cite{He:2019vzf}
\begin{multline}\label{integrals}
\mathscr{I}_{n_1 n_2...n_i \bar{n}_1\bar{n}_2...\bar{n}_j}(w,\bar{w},z_{k_1},z_{k_2},...,z_{k_i},\bar{z}_{l_1},\bar{z}_{l_2},...,\bar{z}_{l_j})=\int \td^2z\frac{(z-w)^2}{(\bar{z}-\bar{w})^2}(\gamma^{z\bar{z}[0]})^2\\
\times\frac{1}{(z-z_{k_1})^{n_1}(z-z_{k_2})^{n_2}...(z-z_{k_i})^{n_i}(\bar{z}-\bar{z}_{l_1})^{\bar{n}_1} (\bar{z}-\bar{z}_{l_2})^{\bar{n}_2}...(\bar{z}-\bar{z}_{l_j})^{\bar{n}_j} },
\end{multline}
where $n_i$ or $\bar{n}_j$ can be any integer. At $\lambda^2$ order, the soft factor is
\begin{multline}
{S}^{(1)-}_{[2]}(\vec{w}|\vec{z}_k)=\frac{2}{\pi}\lambda^{2}
\Bigg\{\sum_{k,l,j}\bigg[\mathscr{I}_{112}(w,\bar{w},z_k,z_l,\bar{z}_j) \hat{\bar{h}}_j +
\frac12\mathscr{I}_{111}(w,\bar{w},z_k,z_l,\bar{z}_j) (\Gamma^{\bar{z}_j}_{\bar{z}_j\bar{z}_j}\hat{\Delta}_j+2\partial_{\bar{z}_j})\bigg]\partial_{z_l}\partial_{z_k}\\
+8\sum_{k,l,j}\bigg[\mathscr{I}_{212}(w,\bar{w},z_k,z_l,\bar{z}_j) \hat{h}_k \hat{\bar{h}}_j + \frac12 \mathscr{I}_{211}(w,\bar{w},z_k,z_l,\bar{z}_j) \hat{h}_k (\Gamma^{\bar{z}_j}_{\bar{z}_j\bar{z}_j}\hat{\Delta}_j+2\partial_{\bar{z}_j})
+ \frac12 \mathscr{I}_{112}(w,\bar{w},z_k,z_l,\bar{z}_j)   \Gamma^{z_k}_{z_kz_k}\hat{\Delta}_k   \hat{\bar{h}}_j\\
+ \frac14 \mathscr{I}_{111}(w,\bar{w},z_k,z_l,\bar{z}_j) \Gamma^{z_k}_{z_kz_k}\hat{\Delta}_k  (\Gamma^{\bar{z}_j}_{\bar{z}_j\bar{z}_j}\hat{\Delta}_j+2\partial_{\bar{z}_j})\bigg]\partial_{z_l}
+4 \sum_{k,l,j} \bigg[\mathscr{I}_{222}(w,\bar{w},z_k,z_l,\bar{z}_j)\hat{h}_k \hat{h}_l \hat{\bar{h}}_j\\
+ \frac12 \mathscr{I}_{221}(w,\bar{w},z_k,z_l,\bar{z}_j) \hat{h}_k \hat{h}_l (\Gamma^{\bar{z}_j}_{\bar{z}_j\bar{z}_j}\hat{\Delta}_j+2\partial_{\bar{z}_j})
+\frac12 \mathscr{I}_{212}(w,\bar{w},z_k,z_l,\bar{z}_j)\hat{h}_k \Gamma^{z_l}_{z_lz_l}\hat{\Delta}_l  \hat{\bar{h}}_j\\
+ \frac14 \mathscr{I}_{211}(w,\bar{w},z_k,z_l,\bar{z}_j) \hat{h}_k \Gamma^{z_l}_{z_lz_l}\hat{\Delta}_l (\Gamma^{\bar{z}_j}_{\bar{z}_j\bar{z}_j}\hat{\Delta}_j+2\partial_{\bar{z}_j})
+\frac12 \mathscr{I}_{122}(w,\bar{w},z_k,z_l,\bar{z}_j)\Gamma^{z_k}_{z_kz_k}\hat{\Delta}_k \hat{h}_l \hat{\bar{h}}_j\\
+\frac14 \mathscr{I}_{121}(w,\bar{w},z_k,z_l,\bar{z}_j)\Gamma^{z_k}_{z_kz_k}\hat{\Delta}_k \hat{h}_l (\Gamma^{\bar{z}_j}_{\bar{z}_j\bar{z}_j}\hat{\Delta}_j+2\partial_{\bar{z}_j})
+\frac14 \mathscr{I}_{112}(w,\bar{w},z_k,z_l,\bar{z}_j)\Gamma^{z_k}_{z_kz_k}\hat{\Delta}_k \Gamma^{z_l}_{z_lz_l}\hat{\Delta}_l \hat{\bar{h}}_j \\
+\frac18 \mathscr{I}_{111}(w,\bar{w},z_k,z_l,\bar{z}_j)\Gamma^{z_k}_{z_kz_k}\hat{\Delta}_k \Gamma^{z_l}_{z_lz_l}\hat{\Delta}_l (\Gamma^{\bar{z}_j}_{\bar{z}_j\bar{z}_j}\hat{\Delta}_j+2\partial_{\bar{z}_j})\bigg]\\
+4\sum_{k,l}\bigg[\mathscr{I}_{32}(w,\bar{w},z_k,\bar{z}_l) \hat{\bar{h}}_l + \frac12 \mathscr{I}_{31}(w,\bar{w},z_k,\bar{z}_l)(\hat{\Delta}_l\Gamma^{\bar{z}_l}_{\bar{z}_l\bar{z}_l}+2\partial_{\bar{z}_l})
-\mathscr{I}_{111}(w,\bar{w},z_k,z_l,\bar{z}_l)\hat{\Delta}_l\gamma^{[0]}_{z_l\bar{z}_l}\bigg]\partial_{z_k}\\
+4\sum_{k,l}\bigg[2\mathscr{I}_{42}(w,\bar{w},z_k,\bar{z}_l)\hat{h}_k\hat{\bar{h}}_l + \mathscr{I}_{41}(w,\bar{w},z_k,\bar{z}_l)\hat{h}_k\Gamma^{z_l}_{z_lz_l}\hat{\Delta}_l
- \mathscr{I}_{211}(w,\bar{w},z_k,z_l,\bar{z}_l)\hat{h}_k \hat{\Delta}_l \gamma^{[0]}_{z_l\bar{z}_l}
+2\mathscr{I}_{41}(w,\bar{w},z_k,\bar{z}_l)\hat{h}_k \p_{\bz_l}\\
+ \frac12 \mathscr{I}_{31}(w,\bar{w},z_k,\bar{z}_l)\Gamma^{z_k}_{z_kz_k}\hat{\Delta}_k \p_{\bz_l} + \frac14 \mathscr{I}_{21}(w,\bar{w},z_k,\bar{z}_l)(\Gamma^{z_k}_{z_kz_k})^2 \hat{\Delta}_k \p_{\bz_l}
+\frac14 \mathscr{I}_{31}(w,\bar{w},z_k,\bar{z}_l)  \Gamma^{z_k}_{z_kz_k}\hat{\Delta}_k\Gamma^{\bar{z}_l}_{\bar{z}_l\bar{z}_l}\hat{\Delta}_l\\
 + \frac18  \mathscr{I}_{21}(w,\bar{w},z_k,\bar{z}_l) (\Gamma^{z_k}_{z_kz_k})^2\hat{\Delta}_k\Gamma^{\bar{z}_l}_{\bar{z}_l\bar{z}_l}\hat{\Delta}_l
+\frac12 \mathscr{I}_{32}(w,\bar{w},z_k,\bar{z}_l)\hat{\Delta}_k\hat{\bar{h}}_l\Gamma^{z_k}_{z_kz_k}+\frac14 \mathscr{I}_{22}(w,\bar{w},z_k,\bar{z}_l)\hat{\Delta}_k\hat{\bar{h}}_l(\Gamma^{z_k}_{z_kz_k})^2\\
-\frac12 \mathscr{I}_{111}(w,\bar{w},z_k,z_l,\bar{z}_l)\hat{\Delta}_k\hat{\Delta}_l\gamma^{[0]}_{z_l\bar{z}_l}\Gamma^{z_k}_{z_kz_k}\bigg]
-2\sum_{k}\Big[\mathscr{I}_{31}(w,\bar{w},z_k,\bar{z}_k) \hat{\Delta}_k \gamma^{[0]}_{z_k\bar{z}_k} +\mathscr{I}_{21}(w,\bar{w},z_k,\bar{z}_k) \Gamma^{z_k}_{z_kz_k} \hat{\Delta}_k \gamma^{[0]}_{z_k\bar{z}_k}\Big]\Bigg\}.
\end{multline}
The soft factor will involve the following integrals
\begin{equation}\label{integral}
    \mathscr{I}_{n_kn_l\bar{n}_j}(w,\bar{w},z_k,z_l,\bar{z}_j)=\int \td^2z\frac{(z-w)^2}{(\bar{z}-\bar{w})^2}\frac{(\gamma^{z\bar{z}[0]})^2}{(z-z_k)^{n_k}(z-z_l)^{n_l}(\bar{z}-\bar{z}_j)^{\bar{n}_j}},
\end{equation}
which can be calculated separately in the following 4 conditions
\begin{equation}\label{eq:Integral scr}
    \begin{split}
        &k\neq l\neq j:\ \mathscr{I}_{n_kn_l\bar{n}_j}=\int \td^2z\frac{(z-w)^2}{(\bar{z}-\bar{w})^2}\frac{(\gamma^{z\bar{z}[0]})^2}{(z-z_k)^{n_k}(z-z_l)^{n_l}(\bar{z}-\bar{z}_j)^{\bar{n}_j}};\\
        &k\neq l,\ k=j:\ \mathscr{I}_{n_kn_l\bar{n}_k}=\int \td^2z\frac{(z-w)^2}{(\bar{z}-\bar{w})^2}\frac{(\gamma^{z\bar{z}[0]})^2}{(z-z_k)^{n_k}(\bar{z}-\bar{z}_k)^{\bar{n}_k}(z-z_l)^{n_l}};\\
        &k=l\neq j:\ \mathscr{I}_{n_k\bar{n}_j}=\int \td^2z\frac{(z-w)^2}{(\bar{z}-\bar{w})^2}\frac{(\gamma^{z\bar{z}[0]})^2}{(z-z_k)^{n_k}(\bar{z}-\bar{z}_j)^{\bar{n}_j}};\\
        &k=l=j:\ \mathscr{I}_{n_k\bar{n}_k}=\int \td^2z\frac{(z-w)^2}{(\bar{z}-\bar{w})^2}\frac{(\gamma^{z\bar{z}[0]})^2}{(z-z_k)^{n_k}(\bar{z}-\bar{z}_k)^{\bar{n}_k}}.
    \end{split}
\end{equation}
The expressions in \eqref{eq:Integral scr} can be written as a combination of the following 4 integrals
\begin{equation}\label{eq:scri 1-4}
    \begin{split}
        &\mathscr{I}_1(w,\bar{w},z_k,\bar{z}_k)=\int \td^2z\frac{(z-w)^2(1+z\bar{z})^4}{(\bar{z}-\bar{w})|z-z_k|^2}\\
        &\mathscr{I}_2(w,z_k,\bar{z}_j)=\int \td^2z\frac{(z-w)^2(1+z\bar{z})^4}{(z-z_k)(\bar{z}-\bar{z}_j)},\\
        &\mathscr{I}_3(w,\bar{w},z_k)=\int \td^2z\frac{(z-w)^2(1+z\bar{z})^4}{(z-z_k)(\bar{z}-\bar{w})},\\
        &\mathscr{I}_4(w,z_k,\bar{z}_k)=\int \td^2z\frac{(z-w)^2(1+z\bar{z})^4}{|z-z_k|^2}.
    \end{split}
\end{equation}
For $k=l=j$
\begin{equation}\label{eq:nknkbar}
    \begin{split}
        \mathscr{I}_{n_k\bar{n}_k}&=\frac{\partial_{z_k}^{n_k-1}\partial_{\bar{z}_k}^{\bar{n}_k}\partial_{\bar{w}}}{(n_k-1)!(\bar{n}_k-1)!}\int \td^2z \frac{(z-w)^2}{\bar{z}-\bar{w}}\frac{(\gamma^{z\bar{z}[0]})^2}{|z-z_k|^2}\\
        &=\frac{\partial_{z_k}^{n_k-1}\partial_{\bar{z}_k}^{\bar{n}_k}\partial_{\bar{w}}}{4(n_k-1)!(\bar{n}_k-1)!}\mathscr{I}_1(w,\bar{w},z_k,\bar{z}_k).
    \end{split}
\end{equation}
For $k=l\neq j$
\begin{equation}
    \begin{split}
        &\mathscr{I}_{n_k\bar{n}_j}=\frac{\partial_{z_k}^{n_k-1}\partial_{\bar{z}_j}^{\bar{n}_j-1}\partial_{\bar{w}}}{(n_k-1)!(\bar{n}_j-1)!}\int \td^2z\frac{(z-w)^2(\gamma^{z\bar{z}[0]})^2}{(z-z_k)(\bar{z}-\bar{w})(\bar{z}-\bar{z}_j)}\\
        &=\frac{\partial_{z_k}^{n_k-1}\partial_{\bar{z}_j}^{\bar{n}_j-1}\partial_{\bar{w}}}{4(n_k-1)!(\bar{n}_j-1)!}\bigg[\frac{1}{ \bar{z}_{jw}}\Big(\mathscr{I}_2(w,z_k,\bar{z}_j)-\mathscr{I}_3(w,\bar{w},z_k)\Big)\bigg],
    \end{split}
\end{equation}
where
\begin{equation}
    z_{jw}=z_j-w,\quad \bar{z}_{jw}=\bz_j-\bar{w}.
\end{equation}
For $k\neq l\neq j$
\begin{equation}
    \mathscr{I}_{n_kn_l\bar{n}_j}=\frac{\partial_{z_k}^{n_k-1}\partial_{z_l}^{n_l-1}\partial_{\bar{z}_j}^{n_j-1}}{(n_k-1)!(n_l-1)!(\bar{n}_j-1)!}\int \td^2z\frac{(z-w)^2}{(\bar{z}-\bar{w})^2}\frac{(\gamma^{z\bar{z}[0]})^2}{(z-z_k)(z-z_l)(\bar{z}-\bar{z}_j)}.
\end{equation}
So we just need to compute
\begin{equation}
\begin{split}
\mathscr{I}_{11\bar{1}}^{k\neq l}=&\int \td^2z\frac{(z-w)^2}{(\bar{z}-\bar{w})^2}\frac{(\gamma^{z\bar{z}[0]})^2}{(z-z_k)(z-z_l)(\bar{z}-\bar{z}_j)}\\
=&\frac{1}{z_{kl}}\Big[\mathscr{I}_{1\bar{1}}(w,\bar{w},z_k,\bar{z}_j)-\mathscr{I}_{1\bar{1}}(w,\bar{w},z_l,\bar{z}_j)\Big],
\end{split}
\end{equation}
where
\begin{equation}
    \begin{split}
        \mathscr{I}_{1\bar{1}}=\int \td^2z\frac{(z-w)^2}{(\bar{z}-\bar{w})^2}\frac{(\gamma^{z\bar{z}[0]})^2}{(z-z_k)(\bar{z}-\bar{z}_j)}=\frac{\partial_{\bar{w}}}{4}\bigg[\frac{1}{ \bar{z}_{jw}}\Big(\mathscr{I}_2(w,z_k,\bar{z}_j)-\mathscr{I}_3(w,\bar{w},z_k)\Big)\bigg].
    \end{split}
\end{equation}
For $k\neq l,\ k=j$
\begin{equation}
    \mathscr{I}_{n_kn_l\bar{n}_k}=\frac{\partial_{z_k}^{n_k-1}\partial_{z_l}^{n_l-1}\partial_{\bar{z}_j}^{n_j-1}}{(n_k-1)!(n_l-1)!(\bar{n}_j-1)!}\int \td^2z\frac{(z-w)^2(\gamma^{z\bar{z}[0]})^2}{(z-z_l)(z-z_k)(\bar{z}-\bar{z}_k)(\bar{z}-\bar{w})^2}.
\end{equation}
So we just need to compute
\begin{equation}
    \begin{split}
        &\mathscr{I}_{11\bar{1}}^{k=l}=\frac{\partial_{\bar{w}}}{4}\int \td^2z\frac{(z-w)^2(1+z\bar{z})^4}{(z-z_l)(z-z_k)(\bar{z}-\bar{z}_k)(\bar{z}-\bar{w})}\\
        &=\frac{\partial_{\bar{w}}}{4}\bigg[\frac{1}{z_{lk} \bar{z}_{kw}}\Big(\mathscr{I}_2(w,z_l,\bar{z}_k)+\mathscr{I}_3(w,\bar{w},\bar{z}_k)-\mathscr{I}_3(w,\bar{w},\bar{z}_l)-\mathscr{I}_4(w,z_k,\bar{z}_k)\Big)\bigg],
    \end{split}
\end{equation}
where $z_{lk}=z_l-z_k$ and $\bar{z}_{kw}=\bz_k-\bar{w}$.

The explicit forms of the integrals \eqref{eq:scri 1-4} on the celestial sphere can be explicitly worked out. The strategy for computing surface integral in the stereographic coordinates $z=\cot\frac{\theta}{2}e^{i\phi}$, where $(\theta,\phi)$ are the usual spherical variables, is as follows. By virtue of the Stocks' theorem, a contour integral along the boundaries of the surface can be written as a surface integral
\begin{equation}\label{eq:int dz P}
\oint dz P=-\iint d^2z \p_{\bar{z}}P,
\end{equation}
or
\begin{equation}\label{eq:int dzbar P}
\oint d\bar{z}P=\iint d^2z \p_z P.
\end{equation}
Here, we normally need to introduce some boundaries on the celestial sphere to remove some regions where the poles of the integrand are located. Otherwise, one can not apply the Stocks' theorem. Since \eqref{eq:int dz P} and \eqref{eq:int dzbar P} are true for any function $P$, we can rewrite a surface integral as
\begin{equation}\label{surface}
\int d^2z=\oint d\bar{z} \int dz=-\oint dz \int d\bar{z}.
\end{equation}
Note that the first integral $\int \td z$ or $\int \td \bz$ is indefinite integral. See also \cite{Callebaut:2019omt,He:2020udl,Donnay:2021wrk} for relevant discussions. The integration constants from those indefinite integral can not change the results as they will not contribute after the contour integrals. When evaluating the second contour integral, the integrand is not a holomorphic or anti-holomorphic function. In this sense, it is not a usual contour integral. But $z$ and $\bz$ are related from the constraint on the boundaries (the contour). Then one can replace $z$($\bz$) by $\bz$($z$) for the integrand. But, as holomorphic or anti-holomorphic function, the integrand can be different on different parts of the boundary. This is another aspect that the second contour integral in \eqref{surface} is not a usual contour integral. The direction of the contour should be anticlockwise on the boundary or respect to the remaining part of surface. For a compact 2D manifold, the contour will be clockwise respect to the part that cut from the manifold for applying the Stocks' theorem. As an example, we will compute the area of unit 2D sphere. The area of a unite 2D sphere in stereographic coordinates is
\be\label{sphere}\begin{split}
S&=-i \iint \td z \td \bz \frac{2}{(1+z\bz)^2}\\
&=-2i\oint_{\p S} \td z \frac{1}{z(1+z\bz)}.
\end{split}\ee
Note that $z=0$ is a pole of the integrand. So we remove that point (the north point on the sphere where $\theta=\pi$) by introducing a cutoff $\theta=\alpha$. Eventually we will take the limit of $\alpha\to\pi$ to recover the area of the full sphere. Hence the boundary in \eqref{sphere} will be the circle $\theta=\alpha$. On the boundary $z\bz=\cot^2\frac{\alpha}{2}$. Then \eqref{sphere} becomes
\be\begin{split}
S&=-2i\sin^2\frac{\alpha}{2} \oint_{\p S}  \frac{\td z}{z}\\
&=-2i\sin^2\frac{\alpha}{2} \oint_{\p S}  \frac{\td e^{i\phi}}{e^{i\phi}}=4\pi\sin^2\frac{\alpha}{2}.
\end{split}\ee
Clearly, we recover the area of the sphere $4\pi$ when taking the limit $\alpha\to\pi$.

\begin{figure}\begin{center}
\begin{tikzpicture}
\node at (0.05,0) {$\times$};
\draw[thick,xshift=2pt,
decoration={ markings,  
      mark=at position 0.05 with {\arrow{latex}},
      mark=at position 0.3 with {\arrow{latex}},
      mark=at position 0.72 with {\arrow{latex}},
      mark=at position 0.95 with {\arrow{latex}}},
      postaction={decorate}]
  (0.4,0.3) -- (3,0.3) arc (5.7:354.3:3.15)node[above]{$\Lambda_4$} -- (0.4,-0.3);
  \draw[thick,xshift=2pt,
decoration={ markings,
      mark=at position 0.25 with {\arrow{latex}},
      mark=at position 0.85 with {\arrow{latex}}},
      postaction={decorate}]
 (0.4,-0.3)arc (323.13:36.87:0.5)node[above]{$\epsilon_4$} ;
\end{tikzpicture}
\caption{The contour of the $\mathscr{I}_{4}$ integral.}
\label{F3}
\end{center}\end{figure}
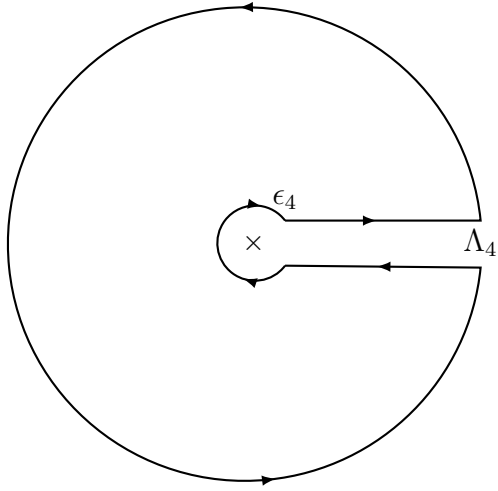

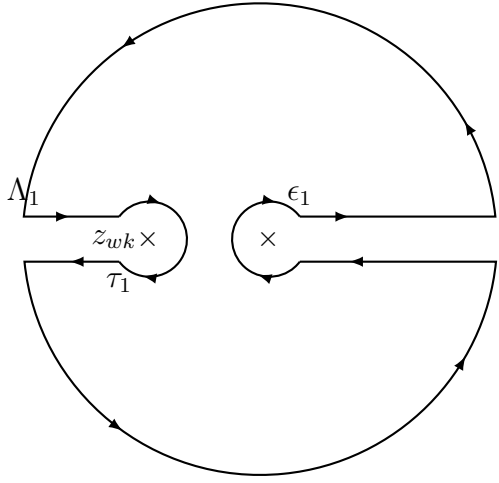
\begin{figure}\begin{center}
\begin{tikzpicture}
\node at (0.05,0) {$\times$};
\node at (-1.55,0) {$\times$};
\node at (-2,0) {$z_{wk}$};
\draw[thick,xshift=2pt,
decoration={ markings,  
      mark=at position 0.05 with {\arrow{latex}},
      mark=at position 0.3 with {\arrow{latex}},
      mark=at position 0.7 with {\arrow{latex}},
      mark=at position 0.95 with {\arrow{latex}}},
      postaction={decorate}]
  (0.4,0.3) -- (3,0.3) arc (5.7:174.3:3.15)node[above]{$\Lambda_1$} -- (-2,0.3);
  \draw[thick,xshift=2pt,
decoration={ markings,  
      mark=at position 0.05 with {\arrow{latex}},
      mark=at position 0.3 with {\arrow{latex}},
      mark=at position 0.7 with {\arrow{latex}},
      mark=at position 0.95 with {\arrow{latex}}},
      postaction={decorate}]
  (-2,-0.3) -- (-3.26,-0.3) arc (185.7:354.3:3.15) -- (0.4,-0.3);
  \draw[thick,xshift=2pt,
decoration={ markings,
      mark=at position 0.25 with {\arrow{latex}},
      mark=at position 0.85 with {\arrow{latex}}},
      postaction={decorate}]
 (0.4,-0.3)arc (323.13:36.87:0.5)node[above]{$\epsilon_1$} ;
   \draw[thick,xshift=2pt,
decoration={ markings,
      mark=at position 0.25 with {\arrow{latex}},
      mark=at position 0.85 with {\arrow{latex}}},
      postaction={decorate}]
 (-2,0.3)arc (143.13:-143.13:0.5)node[below]{$\tau_1$} ;
\end{tikzpicture}
\caption{The contour of the $\mathscr{I}_{1}$ integral.}
\label{F1}
\end{center}\end{figure}

\paragraph{Evaluation of $\mathscr{I}_4$}
We start with the simplest one
\begin{equation}\label{4}
\mathscr{I}_4(w,z_k,\bar{z}_k)=\int \td^2z\frac{(z-w)^2(1+z\bar{z})^4}{|z-z_k|^2}.
\end{equation}
Let us define $z'=z-z_k$ and $\bz'=\bz-\bz_k$. This is nothing but moving the pole $z=z_k$ to the north pole $\theta=\pi$ of the sphere. The integrand is also divergent at $z\to\infty$, the south pole $\theta=0$. In the terms of the new variables $(z',\bz')$, the integral $\mathscr{I}_4(w,z_k,\bar{z}_k)$ becomes
\begin{multline}
\mathscr{I}_4=-\oint \frac{\td z'}{z'}(z' + z_k-w)^2 \bigg[ \frac{1}{4} (z'\bz' +\bz' z_k)^4 + \frac{4}{3} (z'\bz' + \bz' z_k)^3(1+ z' \bz_k + z_k \bz_k)\\
+ 3 (z'\bz' + \bz' z_k)^2(1+z' \bz_k + z_k \bz_k)^2 +
4(z'\bz' + \bz' z_k)(1+z' \bz_k + z_k \bz_k)^3\\
+ (1+z' \bz_k + z_k \bz_k)^4\log \bz'\bigg].
\end{multline}
Since the integrand involves logarithmic function, we need to introduce a branch cut which can not be crossed when choosing the boundary. We introduce the radial coordinate $\rho=\cot \frac{\theta}{2}$. We choose the real axis (the meridian $\phi=0$) as the branch cut. The boundary on the celestial sphere is chosen as demonstrated in Figure \ref{F3}.
Hence the contour integral is divided into four parts
\begin{equation}
\mathscr{I}_{4}=\mathscr{I}_{4a}+\mathscr{I}_{4b}+\mathscr{I}_{4c}+\mathscr{I}_{4d}.
\end{equation}
\begin{multline}
\mathscr{I}_{4a}=\lim_{\epsilon_4\to0}\int_{\epsilon_4} \frac{\td z'}{z'}(z' + z_k-w)^2 \bigg[\frac{4}{z'}(z'+z_k)(1+z' \bz_k + z_k \bz_k)^3 \rho^2\\
+ \frac{3}{{z'}^2} (z'+z_k)^2(1+z' \bz_k + z_k \bz_k)^2 \rho^4 + \frac{4}{3{z'}^3} (z'+z_k)^3(1+z' \bz_k + z_k \bz_k) \rho^6\\
+ \frac{1}{4{z'}^4} (z'+z_k)^4 \rho^8 + (1+z' \bz_k + z_k \bz_k)^4(\log \rho^2 - \log z')\bigg].
\end{multline}
\begin{multline}
\mathscr{I}_{4b}=-\lim_{\Lambda_4\to\infty}\int_{\Lambda_4} \frac{\td z'}{z'}(z' + z_k-w)^2 \bigg[\frac{4}{z'}(z'+z_k)(1+z' \bz_k + z_k \bz_k)^3 \rho^2\\
+ \frac{3}{{z'}^2} (z'+z_k)^2(1+z' \bz_k + z_k \bz_k)^2 \rho^4 + \frac{4}{3{z'}^3} (z'+z_k)^3(1+z' \bz_k + z_k \bz_k) \rho^6\\
+ \frac{1}{4{z'}^4} (z'+z_k)^4 \rho^8 + (1+z' \bz_k + z_k \bz_k)^4(\log \rho^2 - \log z')\bigg].
\end{multline}
\begin{multline}
\mathscr{I}_{4c}=-\int^{\Lambda_4}_{\epsilon_4} \frac{\td \rho}{\rho}(\rho + z_k-w)^2 \bigg[ \frac{1}{4} (\rho^2 +\rho z_k)^4 + \frac{4}{3} (\rho^2 + \rho z_k)^3(1+ \rho \bz_k + z_k \bz_k)\\
+ 3 (\rho^2 + \rho z_k)^2(1+\rho \bz_k + z_k \bz_k)^2 +4(\rho^2 + \rho z_k)(1+\rho \bz_k + z_k \bz_k)^3\\
+ (1+\rho\bz_k + z_k \bz_k)^4\log \rho\bigg].
\end{multline}

\begin{multline}
\mathscr{I}_{4d}=\int^{\Lambda_4}_{\epsilon_4} \frac{\td \rho}{\rho}(\rho + z_k-w)^2 \bigg[ \frac{1}{4} (\rho^2 +\rho z_k)^4 + \frac{4}{3} (\rho^2 + \rho z_k)^3(1+ \rho \bz_k + z_k \bz_k)\\
+ 3 (\rho^2 + \rho z_k)^2(1+\rho \bz_k + z_k \bz_k)^2 +4(\rho^2 + \rho z_k)(1+\rho \bz_k + z_k \bz_k)^3\\
+ (1+\rho\bz_k + z_k \bz_k)^4(\log \rho-2\pi i)\bigg].
\end{multline}
One can easily show that
\begin{equation}
\begin{split}
\mathscr{I}_{4c}+\mathscr{I}_{4d}=&-2\pi i \int^\Lambda_\epsilon \frac{\td \rho}{\rho}(\rho + z_k-w)^2(1+\rho\bz_k + z_k \bz_k)^4\\
=&2\pi i\bigg\{\frac{1}{4} \bz_k^2 \left[3 w^2 \bz_k^2+2 w \bz_k (\bz_k z_k+4)+\bz_k^2 z_k^2+4 \bz_k z_k+6\right] (\rho -w + z_k)^4\\
&+\frac{1}{3} \bz_k \big[w^3 \bz_k^3+w^2 \bz_k^2 (\bz_k z_k+4)+w \bz_k \left(\bz_k^2 z_k^2+4 \bz_k z_k+6\right)+\bz_k^3 z_k^3\\
&+4 \bz_k^2 z_k^2+6 \bz_k z_k+4\big] (\rho -w+z_k)^3+\frac{1}{6} \bz_k^4 (\rho -w+z_k)^6\\
&+\frac{1}{5} \bz_k^3 (3 w \bz_k+\bz_k z_k+4) (\rho -w+z_k)^5+\frac{1}{2} (\bz_k z_k+1)^4 (\rho -w+z_k)^2\\
&-(w-z_k) (\bz_k z_k+1)^4 (\rho -w+z_k)+(w-z_k)^2 (\bz_k z_k+1)^4\log \rho \bigg\}\bigg|_{\Lambda_4}^{\epsilon_4}.\nn
\end{split}
\end{equation}

\begin{equation}
\begin{split}
\mathscr{I}_{4a}=&2\pi i \bigg\{3 \rho ^4 \bigg[w^2 (6 \bz_k^2 z_k^2+6 \bz_k z_k+1)-2 w z_k (10 \bz_k^2 z_k^2+12 \bz_k z_k+3)\\
&+z_k^2 (15 \bz_k^2 z_k^2+20 \bz_k z_k+6)\bigg]+\frac{1}{4} \rho ^8 \left(w^2-10 w z_k+15 z_k^2\right)\\
&+\frac{4}{3} \rho ^6 \bigg[w^2 (4 \bz_k z_k+1)-4 w z_k (5 \bz_k z_k+2)+10 z_k^2 (2 \bz_k z_k+1)\bigg]\\
&+4 \rho ^2 (w-z_k) (\bz_k z_k+1)^2 \left[4 w \bz_k z_k+w-3 z_k (2 \bz_k z_k+1)\right]\\
&+ (w-z_k)^2 (\bz_k z_k+1)^4 \log \rho ^2 \\
&-\frac{1}{6}\rho^6 \bz_k^4+\frac{2}{5} \rho^5 \bz_k^3 (w \bz_k-3 \bz_k z_k-2)\\
&-\frac{1}{4} \rho^4 \bz_k^2 \left[w^2 \bz_k^2-2 w \bz_k (5 \bz_k z_k+4)+15 \bz_k^2 z_k^2+20 \bz_k z_k+6\right]\\
&-\frac{4}{3} \rho^3 \bz_k (\bz_k z_k+1) \left[w^2 \bz_k^2-w \bz_k (5 \bz_k z_k+3)+5 \bz_k^2 z_k^2+5 \bz_k z_k+1\right]\\
&-\frac{1}{2}\rho^2 (\bz_k z_k+1)^2\left[6 w^2 \bz_k^2-4 w \bz_k (5 \bz_k z_k+2)+15 \bz_k^2 z_k^2+10 \bz_k z_k+1\right] \\
&-2 \rho (w-z_k) (\bz_k z_k+1)^3 (2 w \bz_k-3 \bz_k z_k-1)\\
&-(w-z_k)^2 (\bz_k z_k+1)^4(\log \rho + \pi i)\bigg\}.\bigg|_{\rho\to\epsilon_4}\nn
\end{split}
\end{equation}

\begin{equation}
\begin{split}
\mathscr{I}_{4b}=&-2\pi i \bigg\{3 \rho ^4 \bigg[w^2 (6 \bz_k^2 z_k^2+6 \bz_k z_k+1)-2 w z_k (10 \bz_k^2 z_k^2+12 \bz_k z_k+3)\\
&+z_k^2 (15 \bz_k^2 z_k^2+20 \bz_k z_k+6)\bigg]+\frac{1}{4} \rho ^8 \left(w^2-10 w z_k+15 z_k^2\right)\\
&+\frac{4}{3} \rho ^6 \bigg[w^2 (4 \bz_k z_k+1)-4 w z_k (5 \bz_k z_k+2)+10 z_k^2 (2 \bz_k z_k+1)\bigg]\\
&+4 \rho ^2 (w-z_k) (\bz_k z_k+1)^2 \left[4 w \bz_k z_k+w-3 z_k (2 \bz_k z_k+1)\right]\\
&+ (w-z_k)^2 (\bz_k z_k+1)^4 \log \rho ^2 \\
&-\frac{1}{6}\rho^6 \bz_k^4+\frac{2}{5} \rho^5 \bz_k^3 (w \bz_k-3 \bz_k z_k-2)\\
&-\frac{1}{4} \rho^4 \bz_k^2 \left[w^2 \bz_k^2-2 w \bz_k (5 \bz_k z_k+4)+15 \bz_k^2 z_k^2+20 \bz_k z_k+6\right]\\
&-\frac{4}{3} \rho^3 \bz_k (\bz_k z_k+1) \left[w^2 \bz_k^2-w \bz_k (5 \bz_k z_k+3)+5 \bz_k^2 z_k^2+5 \bz_k z_k+1\right]\\
&-\frac{1}{2}\rho^2 (\bz_k z_k+1)^2\left[6 w^2 \bz_k^2-4 w \bz_k (5 \bz_k z_k+2)+15 \bz_k^2 z_k^2+10 \bz_k z_k+1\right] \\
&-2 \rho (w-z_k) (\bz_k z_k+1)^3 (2 w \bz_k-3 \bz_k z_k-1)\\
&-(w-z_k)^2 (\bz_k z_k+1)^4(\log \rho + \pi i)\bigg\}.\bigg|_{\rho\to\Lambda_4}\nn
\end{split}
\end{equation}
Clearly, the terms that are independent of the cutoff in $\mathscr{I}_{4a}$ and $\mathscr{I}_{4b}$ are canceled. Thus there are only divergent terms when we take the limit $\epsilon_4\to0$ and $\Lambda_4\to\infty$. If one properly chooses the relation of $\frac{1}{\epsilon_4}$ and $\Lambda_4$ which will just modify the way of $\epsilon_4$ and $\Lambda_4$ approaching $0$ and $\infty$ respectively, all the divergent terms can be canceled. Hence,
\begin{equation}
\mathscr{I}_4(w,z_k,\bar{z}_k)=0.
\end{equation}
\paragraph{Evaluation of $\mathscr{I}_1$} Next, we consider the first interal
\begin{equation}\label{I1}
\mathscr{I}_1(w,\bar{w},z_k,\bar{z}_k)=\int \td^2z\frac{(z-w)^2(1+z\bar{z})^4}{(\bar{z}-\bar{w})|z-z_k|^2}
\end{equation}
In the $(z',\bar{z}')$ coordinates, the integral $\mathscr{I}_1(w,\bar{w},z_k,\bar{z}_k)$ is reduced to
\begin{multline}
\mathscr{I}_1=-\oint \frac{\td z'}{z'} (z' + z_k-w)^2 \bigg[\bz' (z'+ z_k)^2 \bigg({\bar{w}}^2 (z'+z_k)^2 + 2 \bar{w} (z'+z_k) (z' \bz_k+\bz_k z_k+2)\\
+3 {z'}^2 \bz_k^2 + 2 z' \bz_k (3 \bz_k z_k+4)+3 \bz_k^2 z_k^2 + 8 \bz_k z_k + 6 \bigg)\\
+\frac{1}{2} \bz{'^2} (z'+ z_k)^3 [ \bar{w} (z'+ z_k) + 3 z' \bz_k+3 \bz_k z_k+4] + \frac{1}{3} \bz{'^3} (z'+z_k)^4\\
+\frac{[\bar{w} (z'+z_k)+1]^4 \log (\bz'+\bz_k-\bar{w})}{\bar{w}-\bz_k}-\frac{ (z' \bz_k+\bz_k z_k+1)^4\log \bz'}{\bar{w}-\bz_k}\bigg].
\end{multline}
The integrand is divergent at the north and south pole and the north point is a branch point. There is another branch point $z'=z_{wk}$ where $z_{wk}=w-z_k$. So we need to introduce another branch cut that connecting the point $z_{wk}$ to the north pole. The boundary on the celestial sphere is chosen as in Figure \ref{F1} where we choose the other branch cut the meridian $\phi=\phi_0$. We will introduce new cutoff parameters $\Lambda_2$ and $\epsilon_2$. It is important to notice that the introduction of the cutoff is not from physical aspect, e.g., energy scale etc. The reason for the present computation is to remove the divergence in the integrand which prevents the application of Stocks' theorem for evaluating the surface integral. For each surface integral, one can introduce independent cutoff parameters according to the property of the integrand. In the end, the integral can cover the full surface by taking proper limit of the cutoff parameters. And one can further regularize the results by slightly changing the way that the cutoff parameters approach the proper limit.
We separate the $\mathscr{I}_1$ integral into two parts
\begin{multline}
\mathscr{I'}_{1}=-\oint \frac{\td z'}{z'} (z' + z_k-w)^2 \bigg[\bz' (z'+ z_k)^2 \bigg({\bar{w}}^2 (z'+z_k)^2 + 2 \bar{w} (z'+z_k) (z' \bz_k+\bz_k z_k+2)\\
+3 {z'}^2 \bz_k^2 + 2 z' \bz_k (3 \bz_k z_k+4)+3 \bz_k^2 z_k^2 + 8 \bz_k z_k + 6 \bigg)\\
+\frac{1}{2} \bz{'^2} (z'+ z_k)^3 [ \bar{w} (z'+ z_k) + 3 z' \bz_k+3 \bz_k z_k+4] + \frac{1}{3} \bz{'^3} (z'+z_k)^4\\-\frac{ (z' \bz_k+\bz_k z_k+1)^4\log \bz'}{\bar{w}-\bz_k}\bigg],
\end{multline}
and
\begin{equation}
\mathscr{I''}_{1}=-\oint \frac{\td z'}{z'} (z' + z_k-w)^2
\frac{[\bar{w} (z'+z_k)+1]^4 \log (\bz'+\bz_k-\bar{w})}{\bar{w}-\bz_k}.
\end{equation}
The first part does not have the second branch point $z_{wk}$. So its integral contour can be reduced to the one in Figure \ref{F3}. Thus this part has very similar structure as $\mathscr{I}_4$. One can only obtain divergent terms from this part when taking the limits $\epsilon_1\to0$ and $\Lambda_1\to\infty$. Eventually we will regularize the divergent terms by properly choosing the relation between $\frac{1}{\epsilon_1}$ and $\Lambda_1$. So we will not specify the explicit form of $\mathscr{I'}_{1}$. In the second part, the logarithm will not contribute argument when evaluating the two line integrals between the south and north pole. Hence those two line integrals cancel each other. Then the full contour integral will be reduced to two smaller contour integrals. The first one is
\begin{equation}
\mathscr{I''}_{1a}=\lim_{\epsilon_1\to0}\oint_{\epsilon_1}\frac{\td z'}{z'} (z' + z_k-w)^2
\frac{[\bar{w} (z'+z_k)+1]^4 \log (\bz'+\bz_k-\bar{w})}{\bar{w}-\bz_k}.
\end{equation}
We can massage the logarithmic term as follow
\begin{equation}
\begin{split}
\log (\bz'+\bz_k-\bar{w})=&\log (\bz_k-\bar{w}) + \log \left(1-\frac{\bz'}{\bar{w}-\bz_k}\right)\\
=&\log (\bz_k-\bar{w})-\sum_{k=1}^\infty \frac1k (\frac{\bz'}{\bar{w}-\bz_k})^k,
\end{split}
\end{equation}
where we have applied the relation
\begin{equation}
\text{Li}_1(\frac{\bz'}{\bar{w}-\bz_k})=\sum_{k=1}^\infty \frac1k (\frac{\bz'}{\bar{w}-\bz_k})^k=-\log \left(1-\frac{\bz'}{\bar{w}-\bz_k}\right),\;\text{for}\;\;|\frac{\bz'}{\bar{w}-\bz_k}|<1.
\end{equation}
Then we obtain from the first piece that
\begin{equation}
\mathscr{I''}_{1a}=2\pi i (z_k-w)^2
\frac{(\bar{w}z_k+1)^4 \log (\bz_k-\bar{w})}{\bar{w}-\bz_k}.
\end{equation}
The second one is
\begin{equation}
\mathscr{I''}_{1b}=-\oint \frac{\td z'}{z'} (z' + z_k-w)^2
\frac{[\bar{w} (z'+z_k)+1]^4 \log (\bz'+\bz_k-\bar{w})}{\bar{w}-\bz_k},
\end{equation}

where the contour includes four pieces
\begin{align}
\mathscr{I''}_{1bA}&=-\lim_{\Lambda_1\to\infty}\int_{\Lambda_1} \frac{\td z'}{z'} (z' + z_k-w)^2
\frac{[\bar{w} (z'+z_k)+1]^4 \log (\bz'+\bz_k-\bar{w})}{\bar{w}-\bz_k},\\
\mathscr{I''}_{1bB}&=\lim_{\tau_1\to0}\int_{-\pi}^\pi \frac{\td z'}{z'} (z' + z_k-w)^2
\frac{[\bar{w} (z'+z_k)+1]^4 \log (\bz'+\bz_k-\bar{w})}{\bar{w}-\bz_k},\\
\mathscr{I''}_{1bC}&=\int_{z_{wk}}^{\Lambda_1 e^{i\phi_0}} \frac{\td z'}{z'} (z' + z_k-w)^2 \frac{[\bar{w} (z'+z_k)+1]^4 \log (\bz'+\bz_k-\bar{w})}{\bar{w}-\bz_k},\\
\mathscr{I''}_{1bD}&=-\int_{z_{wk}}^{\Lambda_1 e^{i\phi_0}} \frac{\td z'}{z'} (z' + z_k-w)^2
\frac{[\bar{w} (z'+z_k)+1]^4 [\log (\bz'+\bz_k-\bar{w})+2\pi i]}{\bar{w}-\bz_k},
\end{align}
where $\phi_0=\text{arg}(z_{wk})$ is a constant and $e^{i\phi_0}=\frac{z_{wk}}{|z_{wk}|}$. We introduce new variable $t=z'-z_{wk}$. The first two integrals are reduced to
\begin{align}
\mathscr{I''}_{1bA}&=-\lim_{\Lambda_1\to\infty}\int_{\Lambda_1} \frac{\td t}{t+z_{wk}} t^2
\frac{(\bar{w}t + a)^4 \log \bar t}{\bar{w}-\bz_k},\quad a=1 + \bar{w}(z_{wk}+z_k),\\
\mathscr{I''}_{1bB}&=\lim_{\tau_1\to0}\int_{-\pi}^\pi \frac{\td t}{t+z_{wk}} t^2
\frac{(\bar{w} t + a)^4 \log \bar t}{\bar{w}-\bz_k}=0.
\end{align}
Applying the series expansion
\begin{equation}\frac{1}{t+z_{wk}}=\frac{1}{t} \sum_{n=0}^\infty (\frac{z_{kw}}{t})^n,
\end{equation}
$\mathscr{I''}_{1bA}$ becomes
\begin{equation}\mathscr{I''}_{1bA}=-\frac{1}{\bar{w}-\bz_k}\lim_{\Lambda_1\to\infty}\int_{\Lambda_1} \td t \sum_{n=0}^\infty (\frac{z_{kw}}{t})^n t
(\bar{w}t + a)^4 \log \bar t.
\end{equation}
The terms in this series will be either vanishing or divergent when taking the limit $\Lambda_1\to\infty$. Since we will regularize all the divergent terms, we will not present the explicit formulas here. For the other two integrals, we have
\begin{equation}
\begin{split}
\mathscr{I''}_{1bC}+\mathscr{I''}_{1bD}&=-2\pi i\int_{z_{wk}}^{\Lambda_1 e^{i\phi_0}} \frac{\td z'}{z'} ( z_k-w+z')^2
\frac{[\bar{w} (z_k+z')+1]^4 }{\bar{w}-\bz_k}\\
&=-\frac{2\pi i}{\bar{w}-\bar{z}_k} \left[G_{1b}\left(\Lambda_1\frac{z_{wk}}{|z_{wk}|}\right)-G_{1b}(z_{wk})\right],
\end{split}
\end{equation}
where
\begin{multline}
G_{1b}(z)=\frac{1}{4} \bar{w}^2 \left[3 w^2 \bar{w}^2+2 w \bar{w} (\bar{w} z_k+4)+\bar{w}^2 z_k^2+4 \bar{w} z_k+6\right] (z -w+z_k)^4\\
+\frac{1}{3} \bar{w} \bigg[w^2 \bar{w}^2 (\bar{w} z_k+4)+w^3 \bar{w}^3+w \bar{w} \left(\bar{w}^2 z_k^2+4 \bar{w} z_k+6\right)\\
+\bar{w}^3 z_k^3+4 \bar{w}^2 z_k^2+6 \bar{w} z_k+4\bigg] (z -w+z_k)^3\\
+\frac{1}{6} \bar{w}^4 (z -w+z_k)^6+\frac{1}{5} \bar{w}^3 (3 w \bar{w}+\bar{w} z_k+4) (z -w+z_k)^5\\
+\frac{1}{2} (\bar{w} z_k+1)^4 (z -w+z_k)^2-(w-z_k) (\bar{w} z_k+1)^4 (z -w+z_k)\\
+\log (z ) (w-z_k)^2 (\bar{w} z_k+1)^4.
\end{multline}
Hence,
\begin{equation}
    G_{1b}(z_{wk})=\log (z_{wk}) (w-z_k)^2 (\bar{w} z_k+1)^4.
\end{equation}
Finally we obtain that
\begin{equation}
    \begin{split}
        \mathscr{I}_1&=\frac{2\pi i}{\bar{w}-\bz_k}\left[(z_k-w)^2(\bar{w}z_k+1)^4 \log (\bz_k-\bar{w})+G_{1b}(z_{wk}) \right]\\
        &=\frac{2\pi i}{\bar{w}-\bz_k}(z_k-w)^2(\bar{w}z_k+1)^4 \log (z_{wk} \bz_{kw})
    \end{split}
\end{equation}

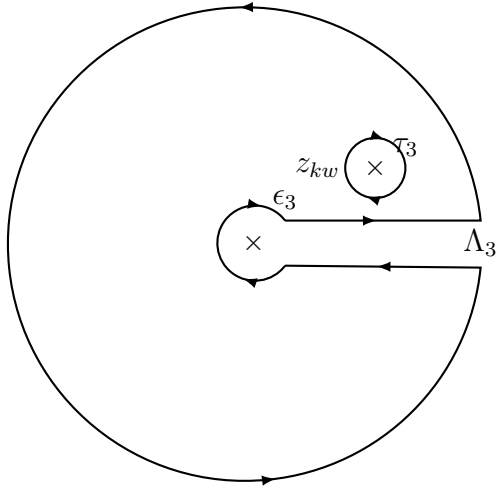
\begin{figure}\begin{center}
\begin{tikzpicture}
\node at (0.05,0) {$\times$};
\node at (1.67,1) {$\times$};
\node at (0.9,1) {$z_{kw}$};
\draw[thick,xshift=2pt,
decoration={ markings,  
      mark=at position 0.05 with {\arrow{latex}},
      mark=at position 0.3 with {\arrow{latex}},
      mark=at position 0.72 with {\arrow{latex}},
      mark=at position 0.95 with {\arrow{latex}}},
      postaction={decorate}]
  (0.4,0.3) -- (3,0.3) arc (5.7:354.3:3.15)node[above]{$\Lambda_3$} -- (0.4,-0.3);
  \draw[thick,xshift=2pt,
decoration={ markings,
      mark=at position 0.25 with {\arrow{latex}},
      mark=at position 0.85 with {\arrow{latex}}},
      postaction={decorate}]
 (0.4,-0.3)arc (323.13:36.87:0.5)node[above]{$\epsilon_3$} ;
   \draw[thick,xshift=2pt,
decoration={ markings,
      mark=at position 0.3 with {\arrow{latex}},
      mark=at position 0.8 with {\arrow{latex}}},
      postaction={decorate}]
 (2,1)arc (360:0:0.4)node[above]{$\tau_3$} ;
\end{tikzpicture}
\caption{The contour of the $\mathscr{I}_{3}$ integral.}
\label{F2}
\end{center}\end{figure}

\paragraph{Evaluation of $\mathscr{I}_3$} We continue with the third integral
\begin{equation}
\mathscr{I}_3(w,\bar{w},z_k)=\int \td^2z\frac{(z-w)^2(1+z\bar{z})^4}{(z-z_k)(\bar{z}-\bar{w})}.
\end{equation}
We define $u=z-w$ in this case. In the new coordinates $(u,\bar{u})$, the integral $\mathscr{I}_3(w,\bar{w},z_k)$ yields
\begin{multline}
\mathscr{I}_3=-\oint \frac{\td u}{u-z_{kw}} u^2 \bigg[\frac{1}{4} {\bar u}^4 (u+w)^4+\frac{4}{3} {\bar u}^3 (u+w)^3 (u {\bar w}+w {\bar w}+1)\\
+3 {\bar u}^2 (u+w)^2 (u {\bar w}+w {\bar w}+1)^2+4 {\bar u} (u+w) (u {\bar w}+w {\bar w}+1)^3\\
+\log ({\bar u}) (u {\bar w}+w {\bar w}+1)^4\bigg].
\end{multline}
The integrand has two branch points $u=0$, $u\to\infty$ and a pole at $u=z_{kw}$. We will choose two separated boundary for this integral on the celestial sphere. One is an infinitesimal circle around the pole $u=z_{kw}$ and the other is to remove the branch points and the branch cut which we choose as the real axis. The contours are shown in Figure \ref{F2}.
We split this integration into two parts
\begin{multline}
\mathscr{I'}_{3}=-\oint \frac{\td u}{u-z_{kw}} u^2 \bigg[\frac{1}{4} {\bar u}^4 (u+w)^4+\frac{4}{3} {\bar u}^3 (u+w)^3 (u {\bar w}+w {\bar w}+1)\\
+3 {\bar u}^2 (u+w)^2 (u {\bar w}+w {\bar w}+1)^2+4 {\bar u} (u+w) (u {\bar w}+w {\bar w}+1)^3\bigg],
\end{multline}
and
\begin{equation}
\mathscr{I''}_{3}=-\oint \frac{\td u}{u-z_{kw}} u^2\log ({\bar u}) (u {\bar w}+w {\bar w}+1)^4.
\end{equation}
The first part is reduced to three loop integrals
\begin{equation}
\mathscr{I'}_{3}=\mathscr{I'}_{3a}+\mathscr{I'}_{3b}+\mathscr{I'}_{3c}.
\end{equation}
\begin{multline}
\mathscr{I'}_{3a}=\lim_{\epsilon_3\to0}\oint_{\epsilon_3} \frac{\td u}{u-z_{kw}} u^2 \bigg[\frac{1}{4} {\bar u}^4 (u+w)^4+\frac{4}{3} {\bar u}^3 (u+w)^3 (u {\bar w}+w {\bar w}+1)\\
+3 {\bar u}^2 (u+w)^2 (u {\bar w}+w {\bar w}+1)^2+4 {\bar u} (u+w) (u {\bar w}+w {\bar w}+1)^3\bigg]=0,
\end{multline}
\begin{equation}
\begin{split}
\mathscr{I'}_{3b}=&\lim_{\tau_3\to0}\oint_{z_{kw}}\frac{\td u}{u-z_{kw}} u^2 \bigg[\frac{1}{4} {\bar u}^4 (u+w)^4+\frac{4}{3} {\bar u}^3 (u+w)^3 (u {\bar w}+w {\bar w}+1)\\
&+3 {\bar u}^2 (u+w)^2 (u {\bar w}+w {\bar w}+1)^2+4 {\bar u} (u+w) (u {\bar w}+w {\bar w}+1)^3\bigg]\\
=&2\pi i z_{kw}^2 \bigg[\frac{1}{4} \bz_{kw}^4 z_k^4+\frac{4}{3} \bz_{kw}^3 z_k^3 (z_k {\bar w}+1)+3 \bz_{kw}^2 z_k^2 (z_k {\bar w}+1)^2\\
& +4 \bz_{kw} z_k (z_k {\bar w}+1)^3\bigg],
\end{split}
\end{equation}
\begin{equation}
\begin{split}
\mathscr{I'}_{3c}=&-\lim_{\Lambda_3\to\infty}\oint_{\Lambda_3} \frac{\td u}{u-z_{kw}} u^2 \bigg[\frac{1}{4} {\bar u}^4 (u+w)^4+\frac{4}{3} {\bar u}^3 (u+w)^3 (u {\bar w}+w {\bar w}+1)\\
&+3 {\bar u}^2 (u+w)^2 (u {\bar w}+w {\bar w}+1)^2+4 {\bar u} (u+w) (u {\bar w}+w {\bar w}+1)^3\bigg]\\
=&\frac{\pi i \Lambda_3^2 }{6}\bigg[8 w^3 {\bar w} \left[\Lambda_3^2 (8 \Lambda_3^2+9)+24 {\bar w}^2 z_{kw}^2+18 (\Lambda_3^2+1) {\bar w} z_{kw}\right]\\
&+12 w^4 {\bar w}^2 \left(3 \Lambda_3^2+4 {\bar w} z_{kw}\right)+6 w^2 \big[\Lambda_3^2  (3 \Lambda_3^4+8 \Lambda_3^2+6 )+48 {\bar w}^3 z_{kw}^3\\
&+36 (\Lambda_3^2+2 ) {\bar w}^2 z_{kw}^2+4  (4 \Lambda_3^4+9 \Lambda_3^2+6 ) {\bar w} z_{kw}\big]
+4 w z_{kw} \big[48 {\bar w}^3 z_{kw}^3\\
&+3  (\Lambda_3^6+4 \Lambda_3^4+6 \Lambda_3^2+4 )+36  (\Lambda_3^2+3 ) {\bar w}^2 z_{kw}^2\\
&+2  (8 \Lambda_3^4+27 \Lambda_3^2+36 ) {\bar w} z_{kw}\big]+z_{kw}^2 \big[3 \Lambda_3^6+16 \Lambda_3^4+36 \Lambda_3^2\\
&+48 {\bar w}^3 z_{kw}^3+36  (\Lambda_3 ^2+4 ) {\bar w}^2 z_{kw}^2+8  (2 \Lambda_3^4+9 \Lambda_3^2+18 ) {\bar w} z_{kw}+48\big]\bigg].
\end{split}
\end{equation}
The second part includes five pieces
\begin{equation}
\mathscr{I''}_{3a}=\lim_{\epsilon_3\to0}\int_{\epsilon_3} \frac{\td u}{u-z_{kw}} u^2\log ({\bar u}) (u {\bar w}+w {\bar w}+1)^4=0.
\end{equation}
\begin{equation}
\mathscr{I''}_{3b}=-\int_{\epsilon_3}^{\Lambda_3} \frac{\td \rho}{\rho-z_{kw}} \rho^2\log (\rho) (\rho {\bar w}+w {\bar w}+1)^4.
\end{equation}

\begin{equation}
\mathscr{I''}_{3c}=\int_{\epsilon_3}^{\Lambda_3} \frac{\td \rho}{\rho-z_{kw}} \rho^2(\log \rho-2\pi i) (\rho {\bar w}+w {\bar w}+1)^4.
\end{equation}
\begin{equation}
\mathscr{I''}_{3d}=-\lim_{\Lambda_3\to\infty}\int_{\Lambda_3} \frac{\td u}{u-z_{kw}} u^2\log ({\bar u}) (u {\bar w}+w {\bar w}+1)^4.
\end{equation}
\begin{equation}
\begin{split}
\mathscr{I''}_{3e}=&\lim_{\tau_3\to0}\oint \frac{\td u}{u-z_{kw}} u^2\log ({\bar u}) (u {\bar w}+w {\bar w}+1)^4\\
=&2\pi i z_{kw}^2\log (\bz_{kw}) (z_{k} {\bar w}+1)^4.
\end{split}
\end{equation}
Similar to the previous case, $\mathscr{I''}_{3d}$ only has divergent terms when the limit $\Lambda_3\to\infty$ is applied.
The remaining two line integrals can be organized as follows

\begin{multline}
\mathscr{I''}_{3b}+\mathscr{I''}_{3c}=-2\pi i\bigg\{\Lambda_3  z_{kw} (w {\bar w}+{\bar w} z_{kw}+1)^4+\frac{1}{2} \Lambda_3^2 (w {\bar w}+{\bar w} z_{kw}+1)^4\\
+\frac{1}{3} \Lambda_3^3 {\bar w} \bigg[4 w^3 {\bar w}^3+6 w^2 {\bar w}^2 ({\bar w} z_{kw}+2)+4 w {\bar w} \left({\bar w}^2 z_{kw}^2+3 {\bar w} z_{kw}+3\right)+{\bar w}^3 z_{kw}^3\\
+4 {\bar w}^2 z_{kw}^2+6 {\bar w} z_{kw}+4\bigg]
+\frac{1}{4} \Lambda_3^4 {\bar w}^2 \bigg[6 w^2 {\bar w}^2+4 w {\bar w} ({\bar w} z_{kw}+3)\\
+{\bar w}^2 z_{kw}^2+4 {\bar w} z_{kw}+6\bigg]
+\frac{1}{5} \Lambda_3^5 {\bar w}^3 (4 w {\bar w}+{\bar w} z_{kw}+4)+\frac{\Lambda_3^6 {\bar w}^4}{6}\\
+z_{kw}^2 (w {\bar w}+{\bar w} z_{kw}+1)^4 \log (\Lambda_3 -z_{kw})  \\
-z_{kw}^2 ({\bar w} z_{k}+1)^4 \log z_{wk}.\bigg\}
\end{multline}
After the regularization, we obtain
\begin{multline}
\mathscr{I}_3(w,\bar{w},z_k)=2\pi i z_{kw}^2 \bigg[\frac{1}{4} \bz_{kw}^4 z_k^4+\frac{4}{3} \bz_{kw}^3 z_k^3 (z_k {\bar w}+1)+3 \bz_{kw}^2 z_k^2 (z_k {\bar w}+1)^2\\
+4 \bz_{kw} z_k (z_k {\bar w}+1)^3+(z_{k} {\bar w}+1)^4 \log \left(\bz_{kw} z_{wk}\right) \bigg].
\end{multline}

\paragraph{Evaluation of $\mathscr{I}_2$} For evaluating the second integral
\begin{equation}\label{I2}
\mathscr{I}_2(w,z_k,\bar{z}_j)=\int \td^2z\frac{(z-w)^2(1+z\bar{z})^4}{(z-z_k)(\bar{z}-\bar{z}_j)},
\end{equation}
we define $v=z-z_j$. Then this integral is reduced to
\begin{multline}
\mathscr{I}_2=-\oint\frac{\td v}{v-z_{kj}} (v+z_{jw})^2\bigg[\frac{1}{4} \bar{v}^4 (v+z_j)^4+\frac{4}{3} \bar{v}^3 (v+z_j)^3 (v \bz_j+\bz_j z_j+1)\\
+3 \bar{v}^2 (v+z_j)^2 (v \bz_j+\bz_j z_j+1)^2+4 \bar{v} (v+z_j) (v \bz_j+\bz_j z_j+1)^3\\
+\log (\bar{v}) (v \bz_j+\bz_j z_j+1)^4\bigg].
\end{multline}
This integral has a very similar form as $\mathscr{I}_3$. We just present regularized results. The details can be consulted from the previous derivation. The final result is
\begin{multline}
\mathscr{I}_2(w,z_k,\bar{z}_j)=2\pi i z_{kw}^2 \bigg[\frac{1}{4} \bz_{kj}^4 z_k^4+\frac{4}{3} \bz_{kj}^3 z_k^3 (z_k {\bar z_j}+1)+3 \bz_{kj}^2 z_k^2 (z_k {\bar z_j}+1)^2\\
+4 \bz_{kj} z_k (z_k {\bar z_j}+1)^3+(z_{k} {\bar z_j}+1)^4 \log \left(\bz_{kj}z_{jk}\right) \bigg].
\end{multline}

\section{Loop correction and T\={T} deformation}
\label{detail2}

In this section, we will choose $\gamma_{z\bar{z}} = 1$, namely the celestial sphere becomes celestial plane, and the 4D metric is still AFS \cite{Barnich:2016lyg,Compere:2016jwb,Ball:2018prg,Compere:2018ylh,Barnich:2021dta}. In this case, the Minkowskian coordinates can be parameterized as \cite{Freidel:2021dfs}
\begin{equation} \label{eq:4dcoordinate}
    \begin{split}
        & x^{\mu} = u \p_{z} \p_{\bar{z}} \hat{q}^{\mu} + r \hat{q}^{\mu} = \frac{1}{\sqrt{2}} \Big( r(1+z\bar{z}) + u, r(z+\bar{z}), -i r(z-\bar{z}), r(1-z\bar{z}) - u \Big), \\
        & \text{celestial plane}: \ x^A = (x^1,x^2) = \frac{r}{\sqrt{2}} \Big( z+\bar{z}, -i (z-\bar{z}) \Big), \quad A = 1,2,
    \end{split}
\end{equation}
where
\begin{equation} \label{eq:Minkq}
    q^{\mu} = \omega \hat{q}^{\mu} = \frac{\omega}{\sqrt{2}} \Big( 1+z\bar{z}, z+\bar{z}, -i(z-\bar{z}), 1-z\bar{z} \Big)
\end{equation}
is the massless momentum parameterized by 4D Minkowskian coordinates. 

\subsection{First order correction of T\={T} deformed correlator}
We will use the standard CFT coordinate $(Z,\Bar{Z})$ to discuss the theory on celestial plane
\begin{equation} \label{eq:Z=rz}
    Z = x_1 + i x_2 = \sqrt{2} R z, \quad \Bar{Z} = x_1 - i x_2 = \sqrt{2} R \Bar{z}.
\end{equation}
Here $R$ should be considered as a dimensionful constant which is to construct the dimensionful coordinates. Hence all the results in \cite{Kapec:2016jld} can be recovered using the coordinates $(Z,\Bar{Z})$. On the celestial plane, one has $R = \text{Constant} \rightarrow \infty$. The $T\bar{T}$ deformed correlator on celestial plane is
\begin{equation}
    \<X_n\>_{[\lambda]} = \left\< \left(1 - \lambda \int d^2 x T_{ZZ}^{[0]} T_{\Bar{Z} \Bar{Z}}^{[0]} + \mathcal{O}(\lambda^2) \right) X_n \right\>.
\end{equation}
The first-order correction of the correlator
\begin{equation} \label{eq:2DXn(1)}
    \<X_n\>_{[\lambda]}^{(1)} = - \lambda \sum_{i \neq j} \left( I_{i \bar{j}}^{2 \bar{2}} h_i \bar{h}_j + I_{i \bar{j}}^{2 \bar{1}} h_i \p_{\bar{Z}_j} + I_{i \bar{j}}^{1 \bar{2}} \bar{h}_j \p_{Z_i} + I_{i \bar{j}}^{1 \bar{1}} \p_{\bar{Z}_j} \p_{Z_i} \right) \<X_n\>
\end{equation}
where we used Ward identity of stress tensor on the plane, and the integrals are as follows
\begin{equation} \label{eq:integrals}
    I_{i \bar{j}}^{r \bar{s}}=\int  \frac{d^2x}{(Z-Z_i)^r (\bar{Z}-\bar{Z}_j)^s} = \frac{\p_{Z_i}^{r-1} \p_{\bar{Z}_j}^{s-1}}{(r-1)! (s-1)!} I_{i \bar{j}}^{1 \bar{1}}, \quad d^2 x = dx^1 \land dx^2 = \frac{i}{2} d^2Z.
\end{equation}
The integrals $I_{i \bar{j}}^{1 \bar{1}}$ can be computed by applying Stokes formula discussed in section \ref{detail1}
\begin{equation}
    I_{i \bar{j}}^{1 \bar{1}} = -\frac{i}{2}\oint dZ \int\frac{d\bar{Z}}{(Z-Z_{ij}) \bar{Z}}=-\frac{i}{2}\oint dZ \frac{\log \bar{Z}}{Z-Z_{ij}}.
\end{equation}
The contour is the same as Fig \ref{F2} and we choose different cut-off parameters for different $ij$, denoted as $\Lambda_{ij}$. Then $I_{i \bar{j}}^{1 \bar{1}}$ can be divided into 4 parts
\begin{equation}
    \begin{split}
        I_{\Lambda_{ij}}=&\frac{1}{2}\int^{2\pi}_0 \Lambda_{ij} e^{i\theta} d\theta \frac{\log\Lambda_{ij}-i\theta}{\Lambda_{ij} e^{i\theta}-Z_{ij}} = \frac{1}{2}\int^{2\pi}_0 d\theta (\log\Lambda_{ij}-i\theta)+\frac{Z_{ij}}{2} \int^{2\pi}_0 d\theta \frac{\log\Lambda_{ij}-i\theta}{\Lambda_{ij} e^{i\theta}-Z_{ij}}\\
        =&\pi \log \Lambda_{ij}-\pi^2 i+0;
    \end{split}
\end{equation}
\begin{equation}
    I_{\epsilon_3}=-\frac{1}{2}\int^{2\pi}_0 \epsilon_3 e^{i\theta} d\theta \frac{\log\epsilon_3-i\theta}{\epsilon_3 e^{i\theta}-Z_{ij}}\rightarrow 0;
\end{equation}
\begin{equation}
    \begin{split}
        I_{\tau_3}=&\frac{i}{2}\oint_{|Z-Z_{ij}|=\tau_3} dZ \frac{\log \left(\bar{Z}_{ij}+\frac{\tau_3^2}{Z-Z_{ij}}\right)}{Z-Z_{ij}} =\frac{i}{2}\oint_{|Z'|=\tau_3} \frac{dZ'}{Z'}\log \left(\bar{Z}_{ij}+\frac{\tau_3^2}{Z'}\right)\\
        =&\frac{i}{2}\oint_{|Z'|=\tau_3} \frac{dZ'}{Z'}\left[\log \bar{Z}_{ij}+\log \left(1+\frac{\tau_3^2}{Z'\bar{Z}_{ij}}\right)\right]= -\pi \log \bar{Z}_{ij}-\frac{i}{2} \sum_{n=1}^{\infty} \frac{1}{n} \oint_0 \frac{dz}{Z^{n+1}}\left(\frac{\tau_3^2}{\bar{Z}_{ji}}\right)^n\\
        =&-\pi \log \bar{Z}_{ij}-0;
    \end{split}
\end{equation}
\begin{equation}
    \begin{split}
        I_{x}=&-\frac{i}{2} \int^{\Lambda_{ij}}_{\epsilon_3} dx \frac{\log x}{x-Z_{ij}}+\frac{i}{2} \int^{\Lambda_{ij}}_{\epsilon_3} dx \frac{\log x-2 \pi i}{x-Z_{ij}} = \pi \int^{\Lambda_{ij}}_{\epsilon_3} \frac{dx}{x-Z_{ij}}=\pi \log \Lambda-\pi \log Z_{ji}.
    \end{split}
\end{equation}
Summing over all terms, one obtains
\begin{equation}
    I_{i \bar{j}}^{1 \bar{1}}= 2\pi \log \Lambda-\pi \log\left(-|Z_{ij}|^2\right)-\pi^2 i = - \pi \log \left( \frac{|Z_{ij}|^2}{\Lambda_{ij}^2} \right).
\end{equation}
Applying the relations in \eqref{eq:Z=rz}, the first-order corrected correlator \eqref{eq:2DXn(1)} is finally obtained as
\begin{equation} \label{eq:deformedXne}
     \<X_n\>_{[\lambda]}^{(1)} = \lambda \pi \sum_{i \neq j} \left[ \frac{h_i}{Z_{ji}} \p_{\bar{Z}_j} + \frac{\bar{h}_j}{\bar{Z}_{ij}} \p_{Z_i} + \left( \log \frac{|Z_{ij}|^2}{\Lambda_{ij}^2} \right) \p_{\bar{Z}_j} \p_{Z_i} \right] \<X_n\>.
\end{equation}
where the anti-symmetric term $\p_{x_i^1} \p_{x_j^2} - \p_{x_i^2} \p_{x_j^1}$ is zero after summation. Note that the second term on the right-hand side can be dropped if we introduce the regularization scheme $|Z-Z_i|>\varepsilon'(\varepsilon' \ll \Lambda_{ij})$ implemented by Cardy \cite{Cardy:2019qao}.

\subsection{Writing one-loop corrected amplitude as 2D correlator}
The inferred divergent part of one-loop corrected gravity amplitude is \cite{Bern:2014oka, Naculich:2011ry}
\begin{equation} \label{eq:1loop}
    \mathcal{A}_{n(1)}^{\text{leading-div}} = G_N \frac{\sigma_n}{\epsilon} \mathcal{A}_n^{(0)}, \quad \sigma_n = \frac{1}{4 \pi} \sum_{i, j = 1}^n (2 q_i \cdot q_j) \log \left( \frac{-2 q_i \cdot q_j}{\mu^2} \right),
\end{equation}
where $\mu$ is the energy scale \cite{Naculich:2011ry}, and $\mathcal{A}_n^{(0)} = \mathcal{A}_n^{(0)} [q_k(\vec{z}_k,\omega_k)]$ is the tree amplitude in momentum space. Here $\vec{z}$ and $\omega$ are used to reparametrize the null momentum. Considering the amplitude in position space $\widetilde{\mathcal{A}}_n^{(0)} (\Vec{z}'_k,r'_k,u'_k)$ which is the Fourier transform of the amplitude in momentum space, the amplitude relation \eqref{eq:1loop} can be rewritten as
\begin{equation} \label{eq:1lloopAn}
    \mathcal{A}_{n(1)}^{\text{leading-div}} = G_N \frac{\sigma_n}{\epsilon} \prod^n_{k=1} \int d^4 x'_k e^{i q_k \cdot x'_k} \widetilde{\mathcal{A}}_n^{(0)} (\Vec{z}'_k,r'_k,u'_k), \quad d^4 x'_k = r^{\prime 2}_k du'_k dr'_k d^2 z'_k .
\end{equation}
Since the scattering data is defined on the null infinity $\mathcal{I}^{\pm}$ ($r' = R \rightarrow \infty$), the amplitude in position space $\widetilde{\mathcal{A}}_n^{(0)} (\Vec{z}'_k,r'_k,u'_k)$ should be determined by the infinity data which is related to Carrollian CFT correlation \cite{Donnay:2022aba, Donnay:2022wvx}. Hence, after inserting a delta function $\delta(r'_k-\infty)$, the amplitude relation \eqref{eq:1lloopAn} is reduced to
\begin{align} \label{eq:An(1)I}
    \mathcal{A}_{n(1)}^{\text{leading-div}} [q_k (\vec{z}_k)] \big|_{\mathcal{I}^{\pm}} & = -  \frac{G_N}{2 \pi \epsilon} \prod^n_{k=1} \int^{\infty}_{-\infty} du'_k e^{- i u'_k \omega_k} \int d^2z'_k e^{- i R \omega_k |z_k-z'_k|^2} \nn\\
    &\hspace{1cm}\times\sum_{i, j = 1}^n  \left( \log \frac{|Z_{ij}|^2}{\Lambda_{ij}^2} \right) \p_{x_i^{\prime \mu}} \p_{x'_{j,\mu}} \widetilde{\mathcal{A}}_n^{(0)} (\Vec{Z}'_k,u'_k) \notag\\
    & = - \frac{G_N}{\pi \epsilon} \sum_{i, j = 1}^n \log \left( \frac{|Z_{ij}|^2}{\Lambda^2} \right) ( \omega_i \omega_j + 2 \p_{Z_i} \p_{\Bar{Z}_{j}} ) \nn \\
    &\hspace{1cm}\times\prod^n_{k=1} \int du'_k e^{- i u'_k \omega_k} \widetilde{\mathcal{A}}_n^{(0)} [\Vec{Z}_k(q_k),u'_k].
\end{align}
In the first equality, we used  \eqref{eq:4dcoordinate} and \eqref{eq:Minkq}, and absorbed $R^2$ into $\widetilde{\mathcal{A}}_n^{(0)} [\Vec{Z}_k(q_k),u'_k]$ to balance the dimension. In the second equality, we used the saddle point approximation ($z'_k = z_k$) following the treatment in \cite{Strominger:2017zoo} for a similar computation. The relation between the length scale $\Lambda_{ij}$ of the celestial plane and the energy scale $\mu$ of the 4D theory is
\begin{equation}
    \sqrt{\omega_i \omega_j} \Lambda_{ij} = R \mu.
\end{equation}
Following \cite{Donnay:2022aba, Donnay:2022wvx},  $\widetilde{\mathcal{A}}_n^{(0)} (\Vec{Z}'_k, u'_k)$ can be interpreted as Carrollian correlator which can be expressed by Carrollian operator $\Phi$
\begin{equation}
    \widetilde{\mathcal{A}}_n^{(0)} (\Vec{Z}'_k, u'_k) = \left\< \prod^n_{k = 1} \Phi_k (u'_k,\vec{Z}'_k) \right\>.
\end{equation}
The correlator relation on the celestial plane is obtained from the amplitude relation \eqref{eq:An(1)I} by Mellin transform which yields
\begin{equation}
    \<X_n\>^{\text{leading-div}}_{(1)} = - \frac{G_N}{\pi \epsilon} \sum_{i, j = 1}^n \log \left( \frac{|Z_{ij}|^2}{\Lambda_{ij}^2} \right) \left( 2 \p_{Z_i} \p_{\Bar{Z}_{j}} - T_i T_j \right) \<X_n\>^{[0]}
\end{equation}
where we used the relation between the Carrollian field and the celestial CFT field \cite{Donnay:2022aba}
\begin{equation} \label{eq:O=Phi}
    \mathcal{O}_{\Delta,s}(\vec{Z}') = \int^{\infty}_0 d\omega \omega^{\Delta-1} \int^{\infty}_{-\infty} du' e^{-i \omega u'} \Phi (u',\vec{Z}') = i^{\Delta} \Gamma (\Delta) \int^{\infty}_{-\infty} du' u^{\prime -\Delta} \Phi (u',\vec{Z}');
\end{equation}
and the action of the operator $T_k$ is
\begin{equation}
    T_k \mathcal{O}_{\Delta_k, s_k} (\Vec{z}_k) = -i \mathcal{O}_{\Delta_k+1, s_k} (\Vec{z}_k).
\end{equation}
Finally, by identifying the $G_N$ and the $\lambda$ as
\begin{equation}
    \lambda = - \frac{2 G_N}{\pi^2 \epsilon},
\end{equation}
the resulting correlator relation \eqref{eq:2Dloop} includes the first-order $T \overline{T}$ deformation result \eqref{eq:deformedXne} which means that the  $T\bar{T}$ deformation captures part of the loop correction information of gravity amplitude.

\bibliography{ref}

\end{document}